\def\year{2018}\relax
\documentclass[letterpaper]{article} %DO NOT CHANGE THIS
\usepackage{aaai18}  %Required
\usepackage{times}  %Required
\usepackage{helvet}  %Required
\usepackage{courier}  %Required
\usepackage{url}  %Required
\usepackage{graphicx}  %Required
\usepackage{booktabs} % For formal tables
\usepackage{epsfig}
\usepackage{graphicx}
\usepackage{amsmath}
\usepackage{amssymb}
\usepackage{bm}

\usepackage{url}
\usepackage{color}
\usepackage[linesnumbered,ruled]{algorithm2e}
\usepackage{caption}
\usepackage{subcaption}

\usepackage{multirow}
\usepackage{siunitx}
\usepackage{xcolor}

\usepackage{array}
\newcolumntype{L}[1]{>{\raggedright\let\newline\\\arraybackslash\hspace{0pt}}m{#1}}
\newcolumntype{C}[1]{>{\centering\let\newline\\\arraybackslash\hspace{0pt}}m{#1}}
\newcolumntype{R}[1]{>{\raggedleft\let\newline\\\arraybackslash\hspace{0pt}}m{#1}}

\DeclareMathOperator*{\argmin}{arg\,min}
\begin{document}
% The file aaai.sty is the style file for AAAI Press 
% proceedings, working notes, and technical reports.
%
\title{The Role of Data-driven Priors in Multi-agent Crowd Trajectory Estimation}
% \author{AAAI Press\\
% Association for the Advancement of Artificial Intelligence\\
% 2275 East Bayshore Road, Suite 160\\
% Palo Alto, California 94303\\
% }
%\maketitle

% % Copyright
% %\setcopyright{none}
% %\setcopyright{acmcopyright}
% %\setcopyright{acmlicensed}
% \setcopyright{rightsretained}
% %\setcopyright{usgov}
% %\setcopyright{usgovmixed}
% %\setcopyright{cagov}
% %\setcopyright{cagovmixed}

% % DOI
% \acmDOI{10.475/123_4}

% % ISBN
% \acmISBN{123-4567-24-567/08/06}

% %Conference
% \acmConference[ICDSC'17]{ACM International Conference on Distributed Smart Camera}{September 5-7, 2017}{Stanford University, CA, USA} 
% \acmYear{2017}
% \copyrightyear{2017}

% \acmPrice{15.00}

% \acmSubmissionID{123-A12-B3}

%\title{Data-driven Priors for Multi-agent Trajectory Interpolation}
%\title{Data-driven Global Optimization for Multi-agent Trajectory Estimation}
%\title{Data-driven Framework for Multi-agent Trajectory Refinements}

%\titlenote{Produces the permission block, and copyright information}
%\subtitle{Extended Abstract}
%\subtitlenote{The full version of the author's guide is available as \texttt{acmart.pdf} document}

%%%%%%%%%%%%%%%%%%%%%%%%%%%%%%%%%%%%%%%%%%%%%%%%%%%%%%%%%%%%%%%%%%%%%%%%%%%%%%%
\author{}
\author{
Gang Qiao$^{1}$, Sejong Yoon$^{2}$, Mubbasir Kapadia$^{1}$, Vladimir Pavlovic$^{1}$ \\
	$^{1}$Rutgers University, $^{2}$The College of New Jersey \\
    \{gq19, mk1353, vladimir\}@cs.rutgers.edu, yoons@tcnj.edu
}
%%%%%%%%%%%%%%%%%%%%%%%%%%%%%%%%%%%%%%%%%%%%%%%%%%%%%%%%%%%%%%%%%%%%%%%%%%%%%%%

%\authornote{Dr.~Trovato insisted his name be first.}
%\orcid{1234-5678-9012}
%\affiliation{%
%  \institution{Computer Science Department, Rutgers University,  The College of New Jersey}
%  \streetaddress{110 Frelinghuysen Road}
%  \city{Piscataway} 
%  \state{New Jersey} 
%  \postcode{08854}
%}
%\email{gq19@cs.rutgers.edu, yoons@tcnj.edu, mk1353@cs.rutgers.edu, vladimir@cs.rutgers.edu}

% The default list of authors is too long for headers}
%\renewcommand{\shortauthors}{B. Trovato et al.}

\maketitle

\begin{abstract}
Trajectory interpolation, the process of filling-in the gaps and removing noise from observed agent trajectories, is an essential task for the motion inference in multi-agent setting.  A desired trajectory interpolation method should be robust to noise, changes in environments or agent densities, while also being yielding realistic group movement behaviors. Such realistic behaviors are, however, challenging to model as they require avoidance of agent-agent or agent-environment collisions and, at the same time, seek computational efficiency. In this paper, we propose a novel framework composed of data-driven priors (local, global or combined) and an efficient optimization strategy for multi-agent trajectory interpolation. The data-driven priors implicitly encode the dependencies of movements of multiple agents and the collision-avoiding desiderata, enabling elimination of costly pairwise collision constraints and resulting in reduced computational complexity and often improved estimation. Various combinations of priors and optimization algorithms are evaluated in comprehensive simulated experiments. Our experimental results reveal important insights, including the significance of the global flow prior and the lesser-than-expected influence of data-driven collision priors.
\end{abstract}

\maketitle

%--------------------------------------------------------
\section{Introduction}
In a multi-agent scenario, a tracking system deployed on an observing, moving or stationary agent (camera), needs to estimate complete trajectories of other moving agents (people, robots, crowds) c.f.,~\cite{bera2014,bera2015}.  However, direct tracker output of the observing agent is often insufficient to reconstruct accurate trajectories because of factors such as noise, the environmental configuration, density of agents in a crowd, or hardware failures. For instance, during the movement of multiple agents in a complex scenario, there are inevitable inter-agent occlusions and agent-obstacle occlusions from the perspective of the observing agent, resulting in observed trajectories that are both noisy and incomplete.

To obtain high-quality complete trajectories, tracking systems typically adopt a multi-step strategy. First, the sensor deployed on the observing agent detects and tracks objects in its neighboring area to obtain \textit{local tracklets}, continuous but short trajectories, of other moving agents. The tracklets of specific agents are then linked together by applying tracklet similarity measurements and re-identification algorithms (e.g., min-cost flow). Next, a trajectory interpolation algorithm is applied to fill in the gaps between the linked tracklets, in a way that the filled portions present desired realistic properties such as containing few potential collisions and saving energy. Finally, trajectory extrapolation could be applied to the reconstructed trajectories so that the observing agent can further plan its own motion.

While several trajectory interpolation approaches have been proposed to-date~\cite{Rodriguez2011,sharma2012,bera2014}, few have proven to work robustly in a general setting,  often exhibiting slow performance and inability to tackle complex scenarios. In practice, many real-time tracking systems typically trade-off accuracy largely for speed~\cite{bera2014}. In addition, few approaches attempt to integrate realistic data-driven priors into the estimation framework except for few very recent attempts in simulation and content generation community~\cite{Bera2016}, resulting in interpolated trajectories that fail to match typical realistic crowd behaviors.

\begin{figure}[t]
 \label{fig1}
 \includegraphics[width=0.48\textwidth]{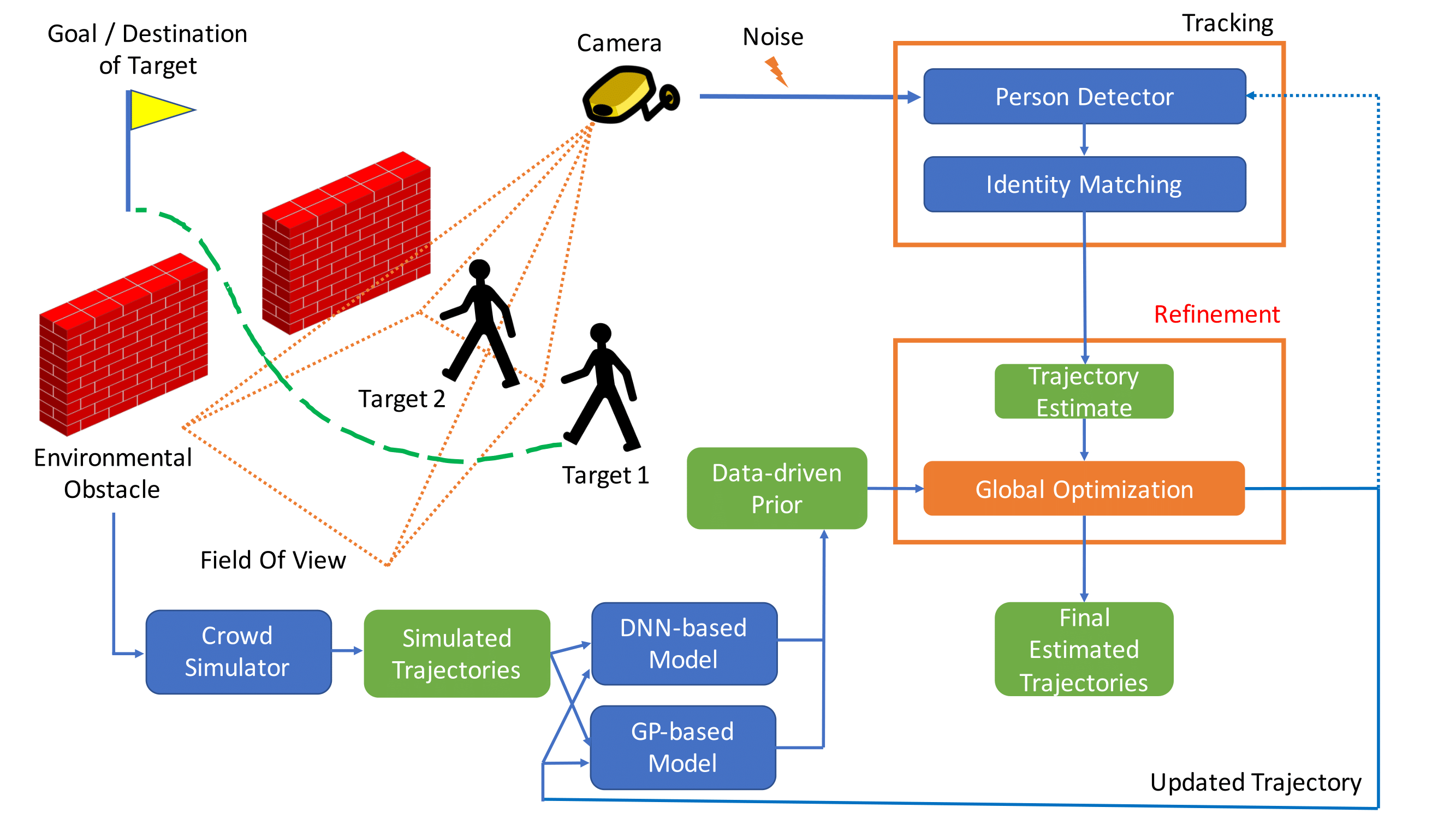}
 \caption{Overview of the proposed framework. A tracker generates noisy and missing trajectories due to occlusions or sensor failure. Our approach takes such observations and applies a global optimization-based trajectory interpolation framework, which incorporates data-driven priors from a model trained on trajectories generated by crowd simulator. Outcome of the optimization (updated trajectories) can be used as new observations to obtain new priors for iteratively improving the tracking performance.}
\vspace{-0.1in}
\end{figure}

In this work we aim to develop a multi-agent trajectory interpolation method of low complexity that results in realistic agent trajectory estimates. To that end, we propose a new multi-agent trajectory interpolation framework composed of a data-driven prior (either local, global or the combination thereof) and an optimization algorithm. The data prior implicitly encodes the movement dependencies of multiple agents and thus can decouple individual agent trajectories, resulting in reduced computational complexity while maintaining or improving the quality of estimation. We evaluate different combinations of prior representations in simulated experiments and demonstrate the essential role of these priors to accomplish the low-complexity, high-accuracy multi-agent trajectory estimation task.

%--------------------------------------------------------
\section{Prior Work}
Modeling, simulation and analysis of crowd motion have attracted significant interest in research community over the past years, e.g.~\cite{musse2010,Ali:2013:MSV:2568128,kapadia2015virtual}. We refer readers to survey papers for important works in this area~\cite{zhan2008,li2015}.  In the following, discuss related work on crowd motion modeling and estimation in context of the trajectory interpolation problem.

\textbf{Crowd Motion and Tracking}. It is common to introduce a motion prior to improve pedestrian tracking accuracy~\cite{hu2008,ali2008,rodriguez2009}. The Social Force model~\cite{PhysRevE.51.4282} has been one particularly popular choice~\cite{pellegrini2009,eth_biwi_00785}, that represents individual dynamic behavior as a combination of different types of forces that characterize both attractive and repulsive effects between pairs of agents or between an agent and an obstacle (an obstacle could be viewed as a static agent with its own shape). Extensions include the multi-target tracking model~\cite{eth_biwi_01014} that incorporates several different priors such as appearance, physical constraints and social behavior of pedestrians. Bera and Manocha~\cite{bera2014,bera2015} introduced a real-time algorithm for trajectory estimation, which works in medium density crowds using particle filtering. Their approach relies on a multi-agent motion model called velocity obstacle~\cite{berg2008} that delimits a space of relative velocity within which collision free behavior is guaranteed for a certain period of time. Similar to these model-based approaches~\cite{pellegrini2009,bera2015}, our approach also relies on a synthetic simulator to provide motion priors for crowd trajectory estimation, but it utilizes the prior as a term within a modified global optimization framework of~\cite{yoon16wacvws}, which is different from the aforementioned approaches. In this way, complex dependencies of movements are implicitly encoded in the prior to decouple trajectories while the optimization provides flexibility of incorporating desired properties in the objective to capture more realistic behaviors. 

\textbf{Crowd Trajectory Estimation}. Alahi et al.~\cite{alahi2014} proposed origin-destination priors to enhance trajectory estimation performance in the context of global optimization-based data-driven methods. They introduced a handcrafted feature called social affinity map (SAM) to capture relative adjoining positions of a number of people within a limited area, by computing the histograms of a number of agents in that region. In contrast, our global optimization framework allows to describe the motion patterns of crowds in a complex scenario within the visibility area of an agent that incorporates much richer information (e.g., the configuration of the environmental obstacles, the shapes and velocities of the agents, rather than a histogram of agent numbers in 2D space) than SAM feature.  Obtaining high-quality simulation data of crowd motion is an important aspect in the context of data-driven crowd trajectory estimation, with many solutions proposed in the graphics community~\cite{kapadia2015virtual}. Since crowd simulation, tracking and trajectory estimation are closely related, crowd simulators~\cite{PhysRevE.51.4282,berg2008} may be utilized as the source of motion priors or training data for motion prior models necessary to improve the tracking accuracy~\cite{journals/cgf/LernerCL07}.

\textbf{Neural Network-based Methods}.  The work in~\cite{long2017} introduced an approach for data-driven collision-minimizing motion planning. In this approach a neural network (NN) is trained from examples of collision-free behaviors, aimed at cloning the Optimal Reciprocal Collision Avoidance (ORCA) policy~\cite{berg2011}, which provides a sufficient condition for avoiding collisions if the agents are not densely packed, otherwise ORCA has to select a reasonable velocity. However, this method can not mimic the movement behavior well without a large amount of training samples. Even with large training sets, emphasizing realistic behaviors such as collision minimization is still difficult, since behavior cloning generates only rough imitations while ignoring some detailed properties without environmental interactions to reinforce those properties. In contrast, our approach can embed various priors, including not only the neural network based local prior, but also globally informed probabilistic velocity priors (see Sec.~\ref{sec:gpflow}) that are critical for proper behaviors, into  a trajectory interpolation framework. This leads to an advantage over either ORCA or neural network-based behavior cloning, resulting in a framework that can handle large portions of missing trajectories. 
% Note that for the local prior, an agent is in its own local coordinate system since each agent is assumed to be self centered without knowing the global world coordinate system. Thus the local observation feature and the predicted velocity from the NN of an agent are all in the local coordinate system of that agent.

%--------------------------------------------------------
\section{Notation}

We use the following notation to define the framework. Typical tracker output at a specific time point is represented with location and a timestamp.  For 2D trackers, the triple $[x_{t}^{i}; y_{t}^{i}; s_{t}^{i}]$ specifies the x-y location and the corresponding timestamp of agent $i$ at video frame (time step) $t$. In this work, we assume a uniform temporal grid with a fixed temporal gap between each consecutive pair of frames of a trajectory. Under this assumption, we can omit the timestamp $s_{t}^{i}$ from the triple.
\newpage
Our goal is to estimate a set of trajectories for $N$ agents $\{ \mathbf{X}^{i} \}_{i=1}^{N}$, where $\mathbf{X}^{i} = \left\{ \mathbf{x}_{t}^{i} \right\}_{t=1}^{T}$ is the desired but unobserved trajectory of agent $i$.  We will be estimating $\{ \mathbf{X}^{i} \}_{i=1}^{N}$ from the corresponding observed fragments of trajectories $\{ \mathbf{O}^{i} \}_{i=1}^{N}$, $ \boldsymbol{O}^i = \left\{ \mathbf{o}_{t}^{i} \right\}_{t=1}^{T}$, with noisy and missing portions. We use the term \emph{tracklet} to refer an observed continuous fragment of a trajectory, and we assume known identification (i.e., correspondences between measurements and estimates) of all tracklets. Let $\Delta t$ denote the sampling period between two consecutive samples $t$ and $t-1$, and $\mathbf{v}_{t}^{i}$ denote the true average velocity of agent $i$ over time interval $[t-1, t]$, thus $\mathbf{x}_{t}^{i} = \mathbf{x}_{t-1}^{i} + {\Delta t} \cdot \mathbf{v}_{t}^{i}$. Index $t = 0$ indicates known initial position of each agent, $\mathbf{x}_{0}^{i}$. 

%--------------------------------------------------------
\section{Data Driven Priors}

We consider two types of data driven priors: local collision avoidance priors and global flow priors.  We next described these priors and how they will be used in the global optimization framework.

%--------------------------------------------------------
\subsection{Local Collision Avoidance Prior}
The goal of utilizing local collision avoidance priors is to effectively replace the collision constraints in~\cite{yoon16wacvws}. We will use the data driven regression approach that aims to estimate the velocity $\mathbf{v}^{i}$ of agent $i$ at space-time point $\mathbf{x}^{i}$ as
\begin{equation}
\label{eq:dnn}
\mathbf{v}^{i}_{t+1}=f_{NN}\left( \left. \mathbf{o}_{t,NN}^{i}, \mathbf{v}^{i}_{t} \right\vert \theta_{NN}\right),
\end{equation}
where $\mathbf{o}_{t,NN}^{i}$ is the local measurement which encodes local visibility of the state space of agent $i$ and $\theta_{NN}$ is the parameter of the model. 

Similar to~\cite{long2017}, the $\mathbf{o}_{t,NN}$ generally includes the following three components: (a) 2D desired velocity: we assume that each agent receives a global velocity guidance signal $\mathbf{\tilde{v}}_{t}^{i}$. This signal is typically a velocity vector pointing toward the goal position of each agent, while disregarding local environment geometry or other agents. (b) Local range/occupancy map: we assume that each agent is equipped with a 360-dimensional distance scanner to collect distances to surface points of other agents and obstacles within a certain range.  Within the distance map relative positions and shapes of the environmental configurations and other agents are explicitly encoded as distances along a 360 degree circularly sampled grid. (c) $360 \times 2$ dimensional velocity map: in addition to aforementioned distances to other agents or obstacles, we also assume the local velocity measurements of neighboring agents or obstacles. 

%--------------------------------------------------------
\subsection{Global Flow Prior\label{sec:gpflow}}
In a multi-agent setting, individual agent movement typically follows a flow pattern that depends on the environment and obstacles, other agents and their density, as well as the global movement goal. This pattern can be encoded in a global flow-field.  While one could obtain it from a path planning algorithms, we instead use a data driven approach to capture the global flow field patterns.  Specifically, we model the flow field as a random walk encoded by a Gaussian Process (GP) prior:
\begin{equation}
\mathbf{x}_{t+\Delta t}^{i} = \mathbf{x}_{t}^{i} + \delta^{i}\left( \mathbf{x}_{t}^{i} \right),
\end{equation}
where 
\begin{equation}
\label{eq:gp}
\delta_{x}^{i}\left(\mathbf{x}_{t}^{i}\right) \sim GP\left( \delta \vert \mathbf{x}_{t}^{i}, s_{t}^{t}, \mathbf{X}_{train}, \theta_{GP}\right).
\end{equation}
$\delta_{x}^{i}(\mathbf{x}_{t}^{i})$ is the "global" velocity of agent $i$ at frame $t$, and $s_{t}^{t}$ denotes the predicted standard deviation of the GP. The GP will now model the data-prior velocity field in $(x,y,s)$ space, similar to \cite{Kim2011-zf}. $\mathbf{X}_{train}$ denotes the trajectory data used to train the GP model and $\theta_{GP}$ the model hyper-parameter. 
% Alternatively, one could use more recent and precise approach to estimate the velocity field, proposed in \cite{Holsclaw2013-ww}.

%--------------------------------------------------------
\section{Embedding Priors into\\Multi-agent Optimization Framework}

The essence of the multi-agent optimization framework is to estimate (interpolate) the trajectories of a set of $N$ agents $\mathbf{X}$ from some set of observations $\mathbf{O}$, given a partially observable environment $\mathbf{Z}$, $Pr\left( \mathbf{X} \vert \mathbf{O}, \mathbf{Z} \right)$.
%$\mathbf{X} = \left\{ \left(\mathbf{x}^{i}, \mathbf{v}^{i}\right) \right\} _{i=1}^{N}$ from some set of observations $\mathbf{O} = \{ \mathbf{o}^{i} \}_{i=1}^{N}$ given a partially observable environment $\mathbf{Z} = \{ \mathbf{z}^{k} \}_{k=1}^{K}$:
%\begin{equation}
%Pr\left( \mathbf{X} \vert \mathbf{O}, \mathbf{Z} \right).
%\end{equation}
%where $k$ denotes the total number of static obstacles in the environment that can be represented as a set of line segments $\mathbf{z}^{k} = [z_{x1}^{k}, z_{x2}^{k}, \cdots; z_{y1}^{k}, z_{y2}^{k}, \cdots]$.
In \cite{yoon16wacvws}, this problem is formulated as a MAP estimation problem by minimizing the following Gibbs energy
\begin{align}\small
\sum_{i} E_{u}^{i}(\mathbf{x}^{i},\mathbf{v}^{i}|\mathbf{o}^{i}) 
	+\sum_{i}\sum_{i\neq j} E_{p}^{i,j}(\mathbf{x}^{i},\mathbf{v}^{i},\mathbf{x}^{j},\mathbf{v}^{j}),
\end{align}
where $E_{u}^{i}$ is the energy term related to an individual agent $i$ (i.e., unary) and $E_{p}^{i,j}$ denotes the pairwise energy term describing dependencies of a pair of agents $(i,j$). The unary term includes energies that model kinematic constraint, maximum velocity constraint as well as the compatibility between the estimated trajectories and the measurements $\mathbf{o}^{i}$. The pairwise term is responsible for avoiding collisions between agents. However, joint optimization of $\mathbf{X}$ is a challenging task, in part because of the existence of the pairwise terms as well as the lack of strong motion priors.  We next describe some of the weak but frequently used priors and then suggest a way to combine our data-driven priors while keeping the computational complexity under control.
%Note that the correspondence of the measurements and the agents is known, hence the term $\mathbf{o}^{i}$ matches $\mathbf{x}^{i}$, $\mathbf{v}^{i}$. In this setting the environmental obstacles are modeled as deterministic static agents. 

%--------------------------------------------------------
\subsection{Existing Unary Priors}
In the work~\cite{yoon16wacvws}, the following three unary energy terms are defined to model individual behavior of an agent: tracker output, kinetic energy, and maximum velocity constraint.  The Tracker Output term seeks to keep the estimated trajectory close to the measured trajectories while taking into account the amount of observation uncertainty $u_{t}^{i}$:
\begin{align}
E_{gt}^{i}( \mathbf{x}^{i} | \mathbf{o}^{i} ) &= \sum_{t} u_{t}^{i} \| \mathbf{x}_{t}^{i} - \mathbf{o}_{t}^{i} \|^{2}.
\end{align}
For instance, $u_t^i = 0$ indicates that the measurement is missing.  The Kinetic Energy term ensures that the total traveled distance is minimized:
\begin{align}
E_{kn}^{i}( \mathbf{x}^{i} ) &= C_{kn} \sum_{t} \| \mathbf{x}_{t}^{i} - \mathbf{x}_{t-1}^{i} \|^{2}.
\end{align}
Parameter $C_{kn}$ can be interpreted as the mass of an agent, which typically could be set as $1$.  Finally, the Maximum Velocity Constraint certifies that each agent's speed not exceed a physically feasible velocity, given as $C_{mv}$:
\begin{align}
E_{mv}^{i}( \mathbf{x}^{i} ) &=
\begin{cases}
0 &\, \text{if } \| \mathbf{v}_{t}^{i} \| \le C_{mv} \\
%0 &\, \text{if } \| \mathbf{x}_{t}^{i} - \mathbf{x}_{t-1}^{i} \| \le C_{mv} \\
\infty &\, \text{otherwise}
\end{cases}, \forall t = 1..T \label{eq:max_speed_contraint}.
\end{align}

%--------------------------------------------------------
\subsection{Combining Local and Global Priors}

The key challenge in the aforementioned multi-agent optimization framework arises from the existence of the pairwise terms, implying coupling trajectories between agents, the highly nonlinear nature of the coupling (collision) constraints and the expensive computations due to this coupling. If such coupling were to be eliminated, the optimization of trajectories for each agent could be solved independently from other agents.  However, their elimination would result in infeasible motion with possibly many collisions.  We therefore propose to replace the computationally costly pairwise terms with stronger global motion and local data-driven collision priors.

Specifically, we propose to modify the unary objectives by augmenting them as follows:
\begin{align}
E_{dg}^{i}( \mathbf{v}^{i} | \mathbf{o}^{i} ) 
	&= \frac{1}{\sigma_{GP}^2\left(\mathbf{o}_{t}^{i},\theta_{GP}\right)}\left\Vert \mathbf{v}_{t+\Delta t}^{i}-\mu_{GP}\left(\mathbf{o}_{t}^{i},\theta_{GP}\right)\right\Vert ^{2} \nonumber \\
    &\quad +\frac{1}{\sigma_{NN}^2}\left\Vert \mathbf{v}_{t+\Delta t}^{i}-f_{NN}\left(\mathbf{o}_{NN}^{i} \vert \theta_{NN}\right)\right\Vert ^{2}\label{eq:new_unary}
\end{align}
Here, $\mu_{GP}$ is the predictive mean and $\sigma_{GP}$ is the predictive standard deviation given by the learned GP and $\sigma_{NN}$ is the standard deviation of the NN regression model.

Integrating all unary terms above, we obtain the final global objective, which we seek to minimize in the estimation process:
% \begin{align}
% \mathbf{\hat{X}} 
% 	&= \argmin_{\mathbf{X}}\sum_{i} E_{u}^{i}(\mathbf{x}^{i},\mathbf{v}^{i}|\mathbf{o}^{i}), \\
% 	&= \argmin_{\mathbf{X}}\sum_{i} \left\{ E_{gt}^{i}( \mathbf{x}^{i} | \mathbf{o}^{i} ) + E_{kn}^{i}( \mathbf{x}^{i} ) \right. \nonumber \\
%     &\quad\quad\quad\quad\quad\quad\quad \left. + E_{mv}^{i}( \mathbf{x}^{i} ) + E_{dg}^{i}( \mathbf{x}^{i} | \mathbf{o}^{i} ) \right\}.
% \label{eq:global_opt}
% \end{align}
\small
\begin{align}
\mathbf{\hat{X}} = \argmin_{\mathbf{X}}\sum_{i} E_{u}^{i}(\mathbf{x}^{i},\mathbf{v}^{i}|\mathbf{o}^{i}) = \argmin_{\mathbf{X}}\sum_{i} E_{gt}^{i}(\mathbf{x}^{i}|\mathbf{o}^{i})\nonumber \\
 + E_{kn}^{i}( \mathbf{x}^{i} ) + E_{mv}^{i}( \mathbf{x}^{i} ) + E_{dg}^{i}( \mathbf{x}^{i} | \mathbf{o}^{i} )
\label{eq:global_opt}
\end{align}
\normalsize

%--------------------------------------------------------
\section{Optimization of the Global Objective}
In the following, we introduce three optimization approaches to solve the optimization problem in Eq.~\ref{eq:global_opt}. In general, our optimization framework is iterative and outlined in Alg.~\ref{alg1}.
\begin{algorithm}[t]
	% required keywords
    \SetKwInOut{Input}{Input}
    \SetKwInOut{Output}{Output}
	% keywords
	\SetKwData{Left}{left}
    \SetKwData{This}{this}
    \SetKwData{Up}{up}
	\SetKwFunction{Union}{Union}
    \SetKwFunction{FindCompress}{FindCompress}
	% input/output
    \Input{$\mathbf{O}$, $(\theta_{NN}, \sigma_{NN})$, $(\theta_{GP}, \mu_{GP}, \sigma_{GP})$, \\$C_{kn}$, $C_{mv}$}
    \Output{$\hat{\mathbf{X}}$}
    % algorithm body
    Initialize $\mathbf{X}$\;
    \Repeat(){$\mathbf{X}$ converges}{
    	Compute NN prior velocities of $\mathbf{X}$ using Eq.~\ref{eq:dnn}\;
        Compute GP prior velocities of $\mathbf{X}$ using Eq.~\ref{eq:gp}\;
        Find $\mathbf{X}$ by minimizing Eq.~\ref{eq:global_opt}\;
    }()
    $\hat{\mathbf{X}} = \mathbf{X}$\;
    % caption & label
	\caption{Proposed Optimization Framework}
    \label{alg1}
%\vspace{-0.2in}    
\end{algorithm}
The iterative nature of our algorithm stems from the coupling between the essential collision-avoiding NN term in Eq.~\ref{eq:dnn} and the solution to the optimization problem. Namely, the NN term's range observations require the knowledge of the agents' locations and velocities, which are the variable we are solving for.  To mitigate this effect, we propose the alternating optimization scheme where the agent's trajectories from a previous iteration are used as the proxies for measurements in Eq.~\ref{eq:dnn}.

The choice of the minimizer in Step-5 of this algorithm is important but less essential. We consider three methods for optimizing optimizing Eq.~\ref{eq:global_opt}: a message-passing algorithm (MPA) of ~\cite{bento13nips,yoon16wacvws}, a general interior point method (IPM), and an unscented Kalman smoother (UKS) that exploits the sequential nature of each (independent) agent's trajectory optimization task while applying a nonlinear Rauch-Tung-Striebel smoother~\cite{Rauch1965}.  Further details of some of the selected approaches are provided in the Supplement.

\section{Experiments}

To evaluate the proposed framework, we consider 6 experimental settings similar to those in~\cite{yoon16wacvws}: 3 different settings of bottlenecks (each contains a challenging egress in evacuation-like scenarios), concentric circle (agents are symmetrically placed along a circle and aim to reach their antipodal positions), two-way and four-way hallways (the environment is divided by two or four building blocks and agents move along the regulated ways).  The configuration of each scenario and the details of training can be found in the supplementary.

%These scenarios were previously introduced in~\cite{steerbench} and are representative settings for studying crowd behaviors. 

% For each scenario, we simulate 3,000 frames of 30-40 agents except for the concentric circle scenario, where there are 20 agents, followed by sub-sampling those frames. 

%The ground truth trajectories are obtained by randomly setting the agents' initial locations and running SteerSuite library driven by social force AI, and are split into training set and testing set. Since the agent density in testing set is the same as the agent density in training set, we call it the basic evaluation. During training, complete ground truth trajectories in the training set are used while in the test phase, each test trajectory contains a missing segment (a challenging portion around 30\% with respect to its total number of temporal points) that needs to be inferred from the initial linear interpolation as the "guess" for the very first complete observations. In this way, when simulating the movements along the inferred complete trajectories in one iteration, prior velocities at those positions could be obtained, thus forming the objective to update the trajectory for the next iteration, illustrated in Fig.\ref{fig:visualization}.

%\textcolor{red}{Did we comment on what is learned within and what across scenarios?  Eg GP is within but NN across.  Also, there are no details whatsoever on the learning of either GP or NN.  Should be mentioned in one paragraph, then stated details are in Supplement.  And this should be here, not in Methods.}
We considered five different prior velocity predictors: Gaussian Process (GP), Neural Network (NN), a linear combination of the NN and GP (LinComb GP+NN), GP-driven NN (GP-fed-NN), and GP-driven ORCA (GP-fed-ORCA). For GP, NN, and LinComb GP+NN, we used the training split of the trajectories to train the data-driven priors we described in Eq.~\ref{eq:dnn}, Eq.~\ref{eq:gp}, and Eq.~\ref{eq:new_unary}, respectively. For GP-fed-NN, we used the outputs of the trained GP (the velocity (mean) and the variance) as two additional input branches to a neural network model, besides the local observation. For GP-fed-ORCA, we used the trained GP's velocity (mean) as the preferred velocity for ORCA~\cite{berg2011}, a common local collision avoiding framework. When training the models, GP was trained within each scenario while the NN was trained across the scenarios, due to distinct model complexities.

\begin{figure}[t]

 \begin{subfigure}{0.22\textwidth}
 \centering
 \includegraphics[width = \textwidth]{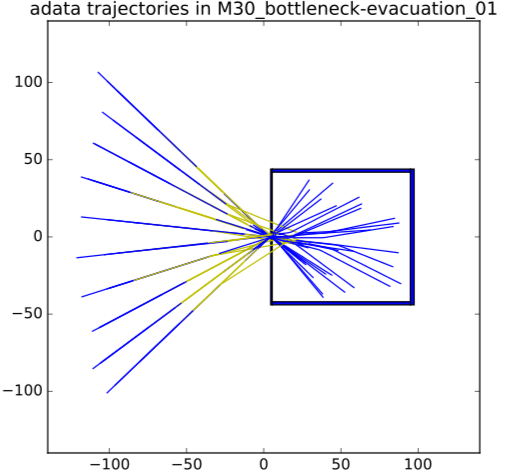}
 \end{subfigure}
 \begin{subfigure}{0.22\textwidth}
 \centering
 \includegraphics[width = \textwidth]{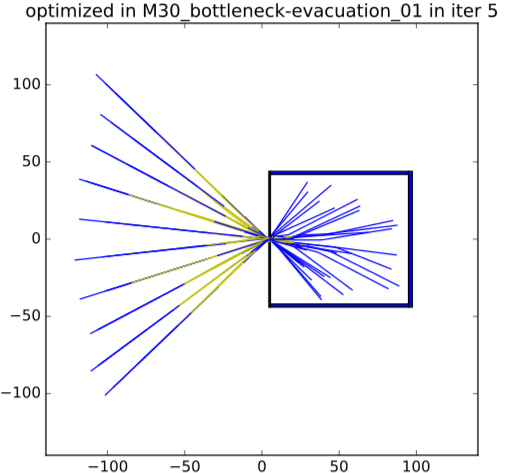}
 \end{subfigure}
 
 \begin{subfigure}{0.22\textwidth}
 \centering
 \includegraphics[width = \textwidth]{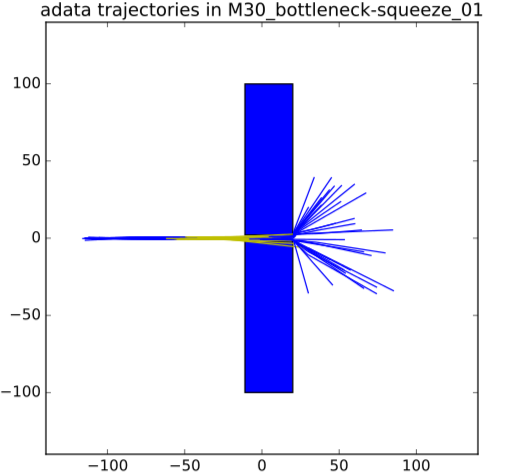}
 \end{subfigure}
 \begin{subfigure}{0.22\textwidth}
 \centering
 \includegraphics[width = \textwidth]{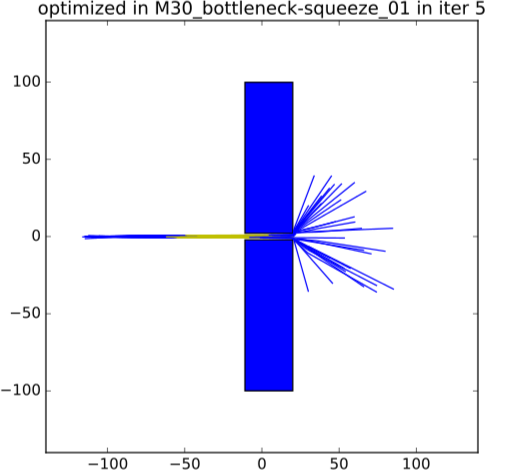}
 \end{subfigure}
 
 \begin{subfigure}{0.21\textwidth}
 \centering
 \includegraphics[width = \textwidth]{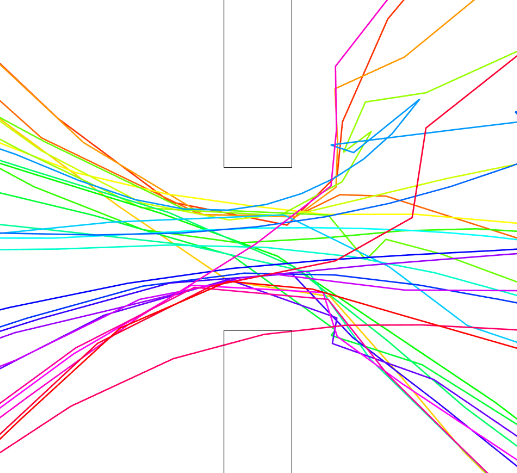}
 \end{subfigure}
 \begin{subfigure}{0.21\textwidth}
 \centering
 \includegraphics[width = \textwidth]{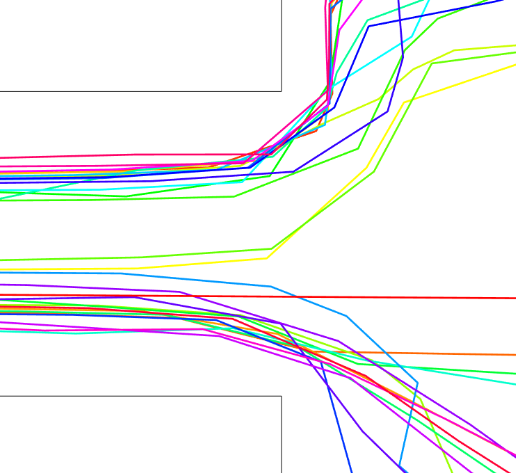}
 \end{subfigure}
 
%  \begin{subfigure}{0.23\textwidth}
%  \centering
%  \includegraphics[width = \textwidth]{images/M30_concentric-circles_01_cropped.png}
%  \end{subfigure}
%  \begin{subfigure}{0.23\textwidth}
%  \centering
%  \includegraphics[width = \textwidth]{images/3D_M30_concentric-circles_01_iter5_cropped.png}
%  \end{subfigure}

\caption{Visualization: the first two rows show the bottleneck evacuation and bottleneck squeeze scenario respectively. In these two rows the left figure denotes the initial linear interpolation and the right figure denotes the optimized trajectories, shown in yellow portions and zoomed in correspondingly in the third row. Best viewed by zooming in.}
% The 4th row is the optimized trajectories in concentric circle scenario, with left 2D and right an additional normalized time axis. Note the rotation behavior in this scenario. 
\label{fig:visualization}
\vspace{-0.2in}
\end{figure}

\textbf{Experimental Setup}. The ground truth trajectories are obtained by running SteerSuite~\cite{steersuite2009} library with social force AI~\cite{PhysRevE.51.4282}, and are split into a training and testing sets. In testing set, each trajectory contains a challenging missing segment around 30\% points, which are initially inferred with linear interpolation. See the illustration in Fig.\ref{fig:visualization}.

%--------------------------------------------------------------------------------------------------
%--------------------------------------------------------------------------------------------------

We consider two evaluation strategies. In the first \textit{basic} strategy, we evaluate our trained models on the test sets from the same agent density setting.  Namely, both the training and the test scenarios contain the matching (identical) number of agents.  In the second \textit{extended }strategy, we consider test scenarios where the agent density varies compared to training setting to evaluate the generalization ability of our approaches.  Details of these evaluations are described below.

The following methods are evaluated: explicit collision avoidance local optimization using message-passing ADMM (MPA)~\cite{yoon16wacvws} that includes pair-wise constraints, our framework with various prior velocity predictor settings (GP, NN, LinComb GP+NN, GP-fed-NN, and GP-fed-ORCA),
%Gaussian Process (GP), Neural Network (NN), a linear combination of NN and GP (LinComb GP+NN), GP-driven NN (GP-fed-NN) and also ORCA driven by GP, 
while the optimizer is either IPM or UKS. We set the parameter as: $u_{t}=1$ if the point is actually observed, otherwise $u_{t}=0$; $C_{kn}=1$, $C_{mv}=2.6m/s$, $\Delta t = 1.5s$ and $\lambda = 1 / (\sigma_{NN}^2 \Delta t^{2}) \approx 108.0$. 

%In contrast to ~\cite{yoon16wacvws} where the point-wise spatial difference between the ground truth trajectories and reconstructed trajectories is the main concern, measured by Root Mean Squared Error, 

We employ three evaluation scores.  The similarity between the ground truth trajectory and reconstructed trajectory is measured with dynamic time warping distance (DTW). We also measure the number of agent-agent collisions and agent-obstacle collisions. An collision occurs when the distance between the centers of two agents is strictly less than the sum of their radii during their continuous movements, and it could be checked by solving a quadratic equation provided with locations of two agents at consecutive time points. The number of collisions is accumulated by counting all collisions along every time step, which means this metric is strict. Note that for simplicity a collision does not change the velocities of involved agents.  We also measure the time-to-completion as a proxy for the computational complexity of each approach.  Indices are measured after the 5th optimization loop.

\begin{figure}[t]
\centering

\begin{subfigure}{0.40\textwidth}
\centering
\includegraphics[width = \textwidth,trim={1cm 1.25cm 1cm 2cm},clip]{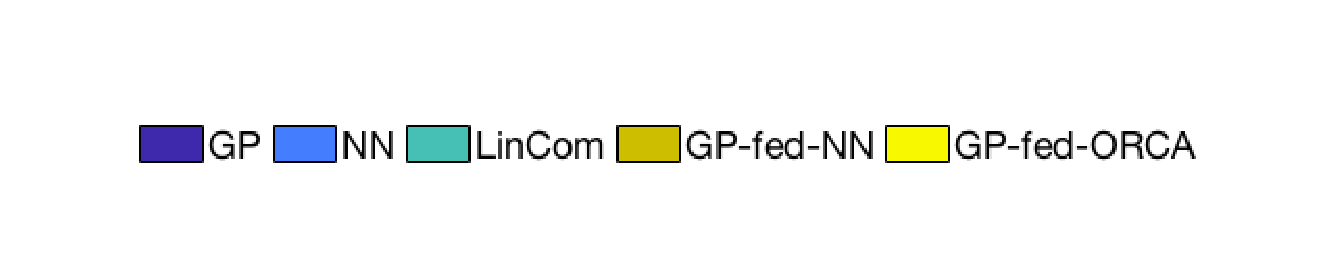}
\end{subfigure}

\begin{subfigure}{0.21\textwidth}
\centering
\includegraphics[width = \textwidth]{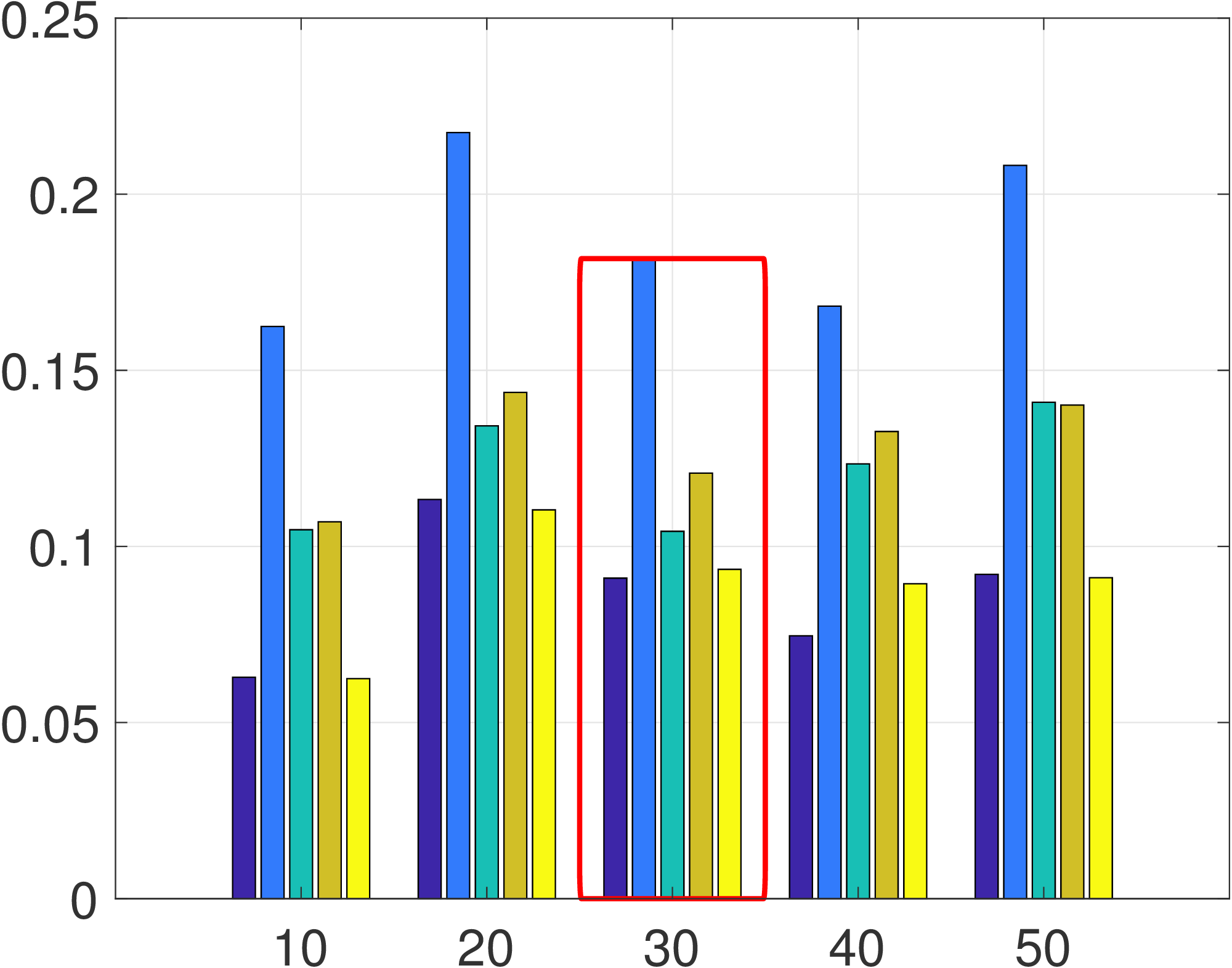}
\caption{bottleneck evacuation}
\end{subfigure}
\begin{subfigure}{0.21\textwidth}
\centering
\includegraphics[width = \textwidth]{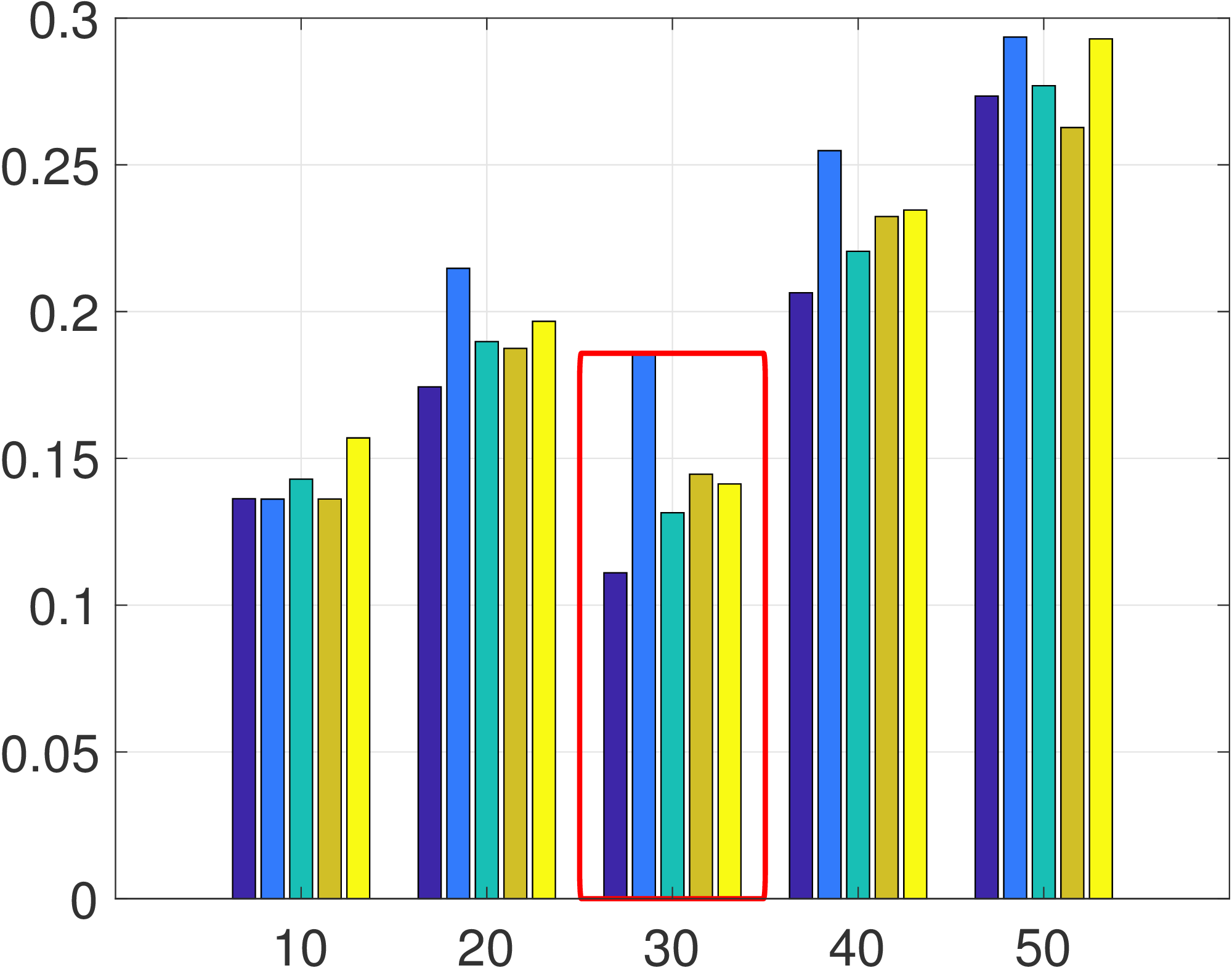}
\caption{bottleneck evacuation 2}
\end{subfigure}

\begin{subfigure}{0.21\textwidth}
\centering
\includegraphics[width = \textwidth]{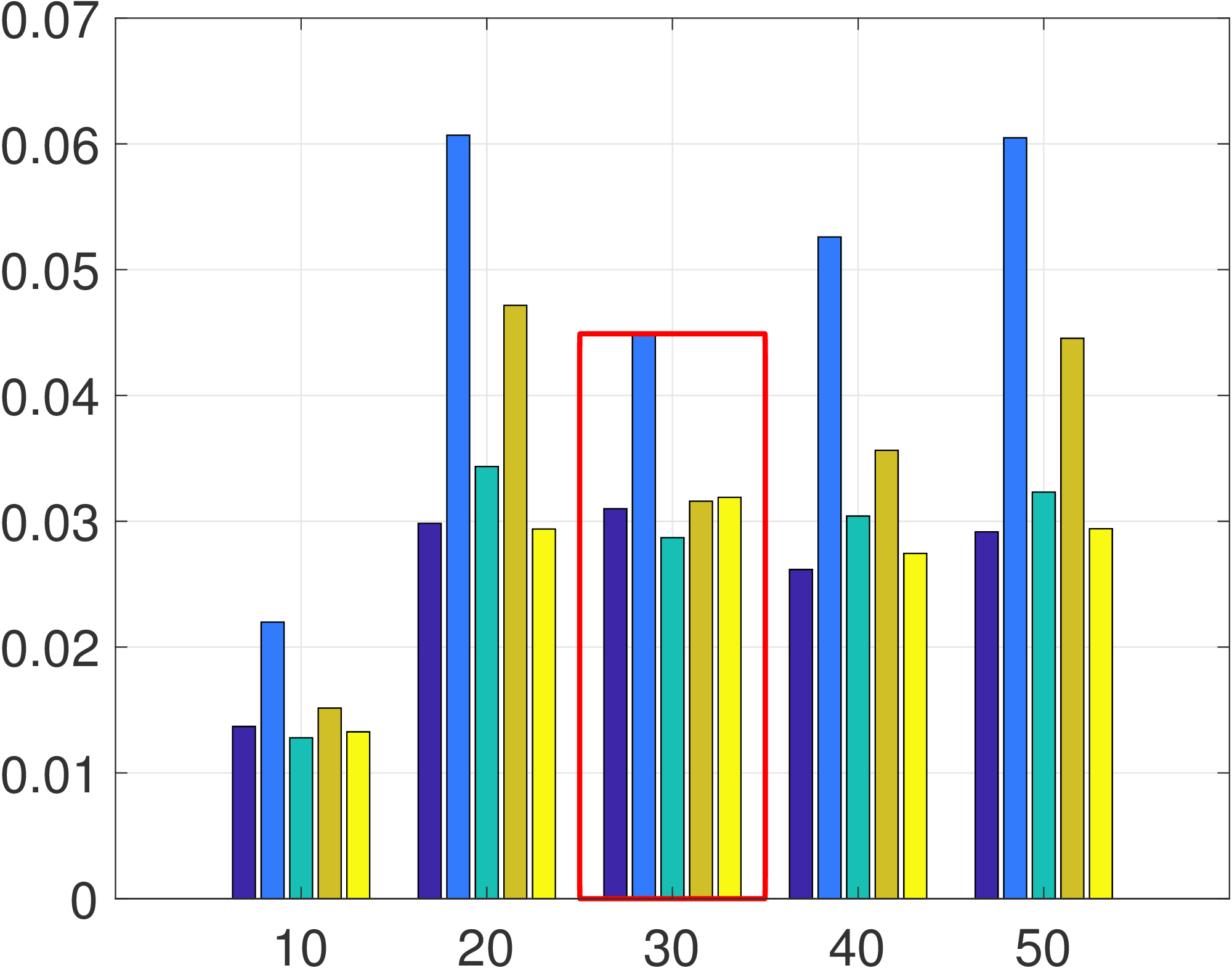}
\caption{bottleneck squeeze}
\end{subfigure}
\begin{subfigure}{0.21\textwidth}
\centering
\includegraphics[width = \textwidth]{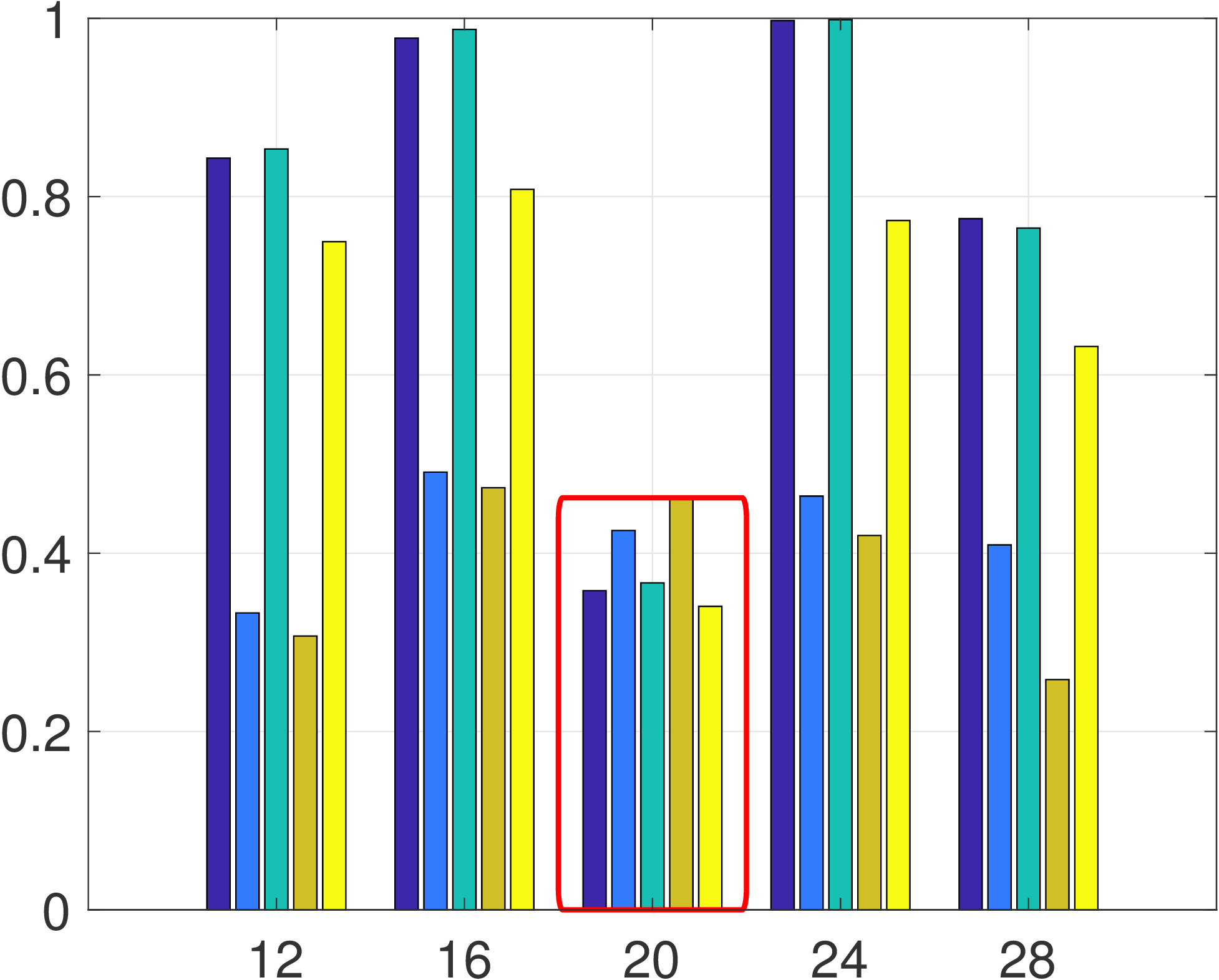}
\caption{concentric circles}
\end{subfigure}

\begin{subfigure}{0.21\textwidth}
\centering
\includegraphics[width = \textwidth]{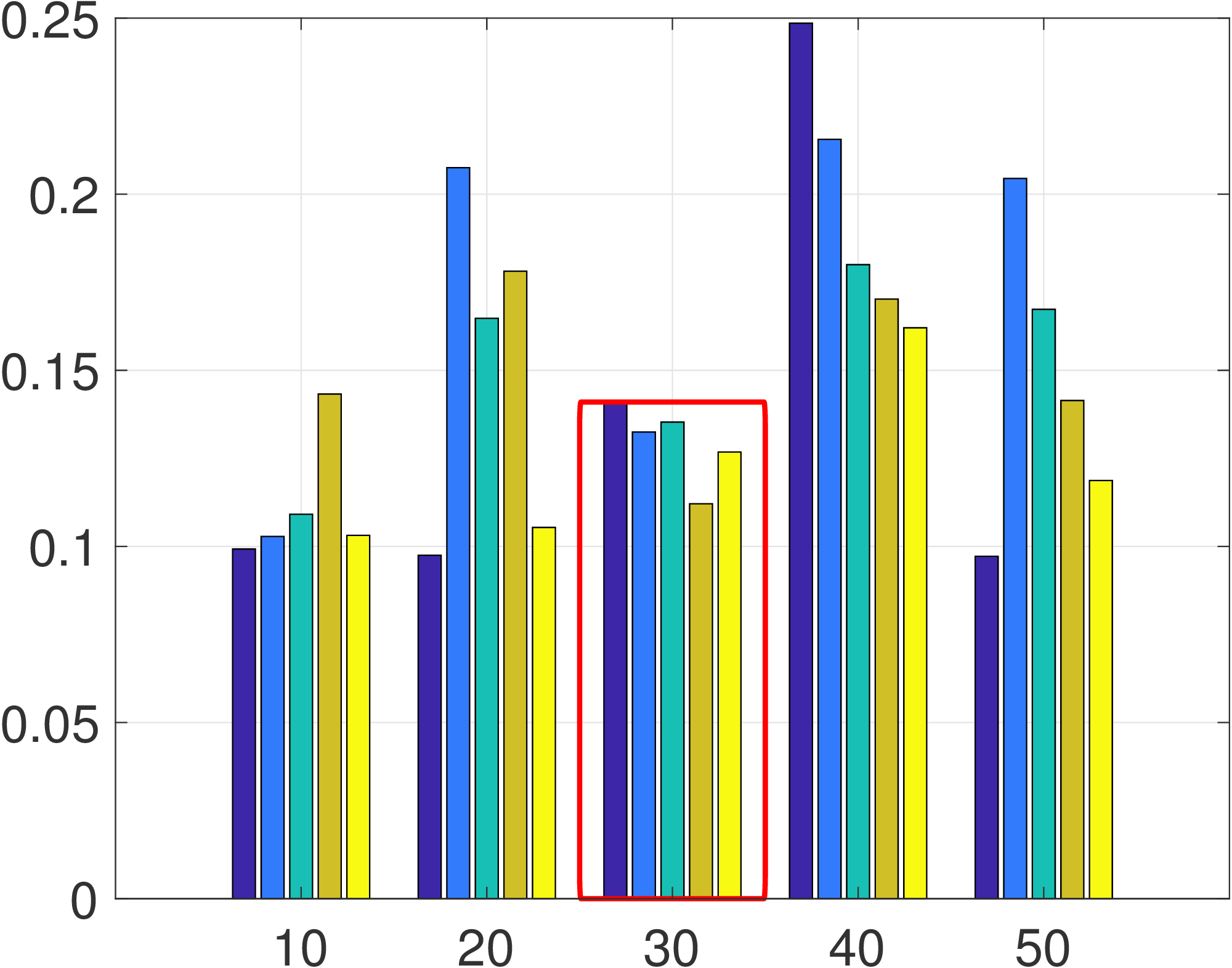}
\caption{hallway-two-way}
\end{subfigure}
\begin{subfigure}{0.21\textwidth}
\centering
\includegraphics[width = \textwidth]{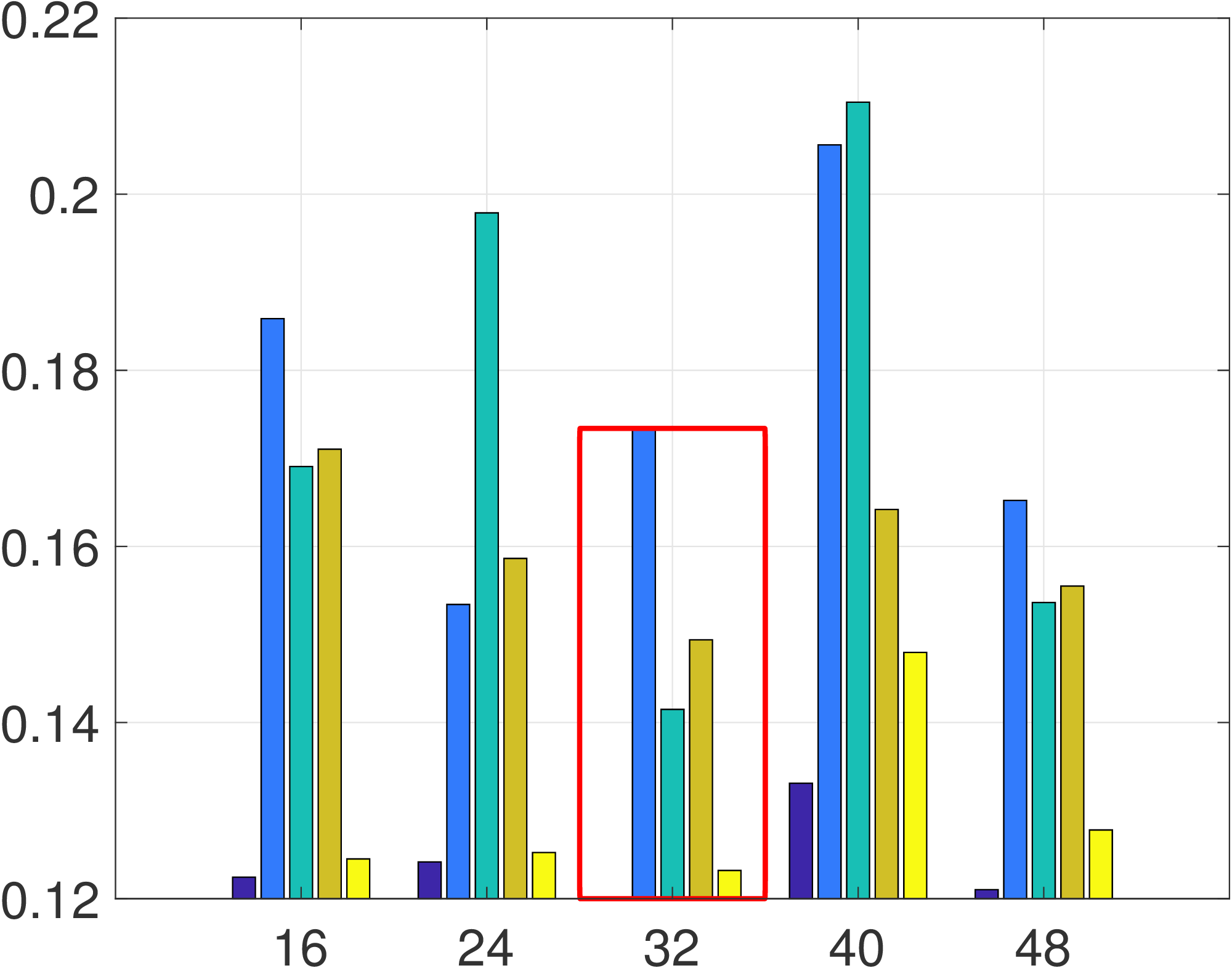}
\caption{hallway-four-way}
\end{subfigure}

\caption{DTW measurements over 6 scenarios using UKS optimization for extended evaluation. Horizontal axis denotes the number of agents (density). Bounding box indicates the case where training and test densities are the same.}
\label{fig:DTW_varying_density}
\vspace{-0.25in}
\end{figure}

\begin{figure}[t]
\centering

\begin{subfigure}{0.40\textwidth}
\centering
\includegraphics[width = \textwidth,trim={1cm 1.25cm 1cm 2cm},clip]{images/legend.png}
\end{subfigure}

\begin{subfigure}{0.21\textwidth}
\centering
\includegraphics[width = \textwidth]{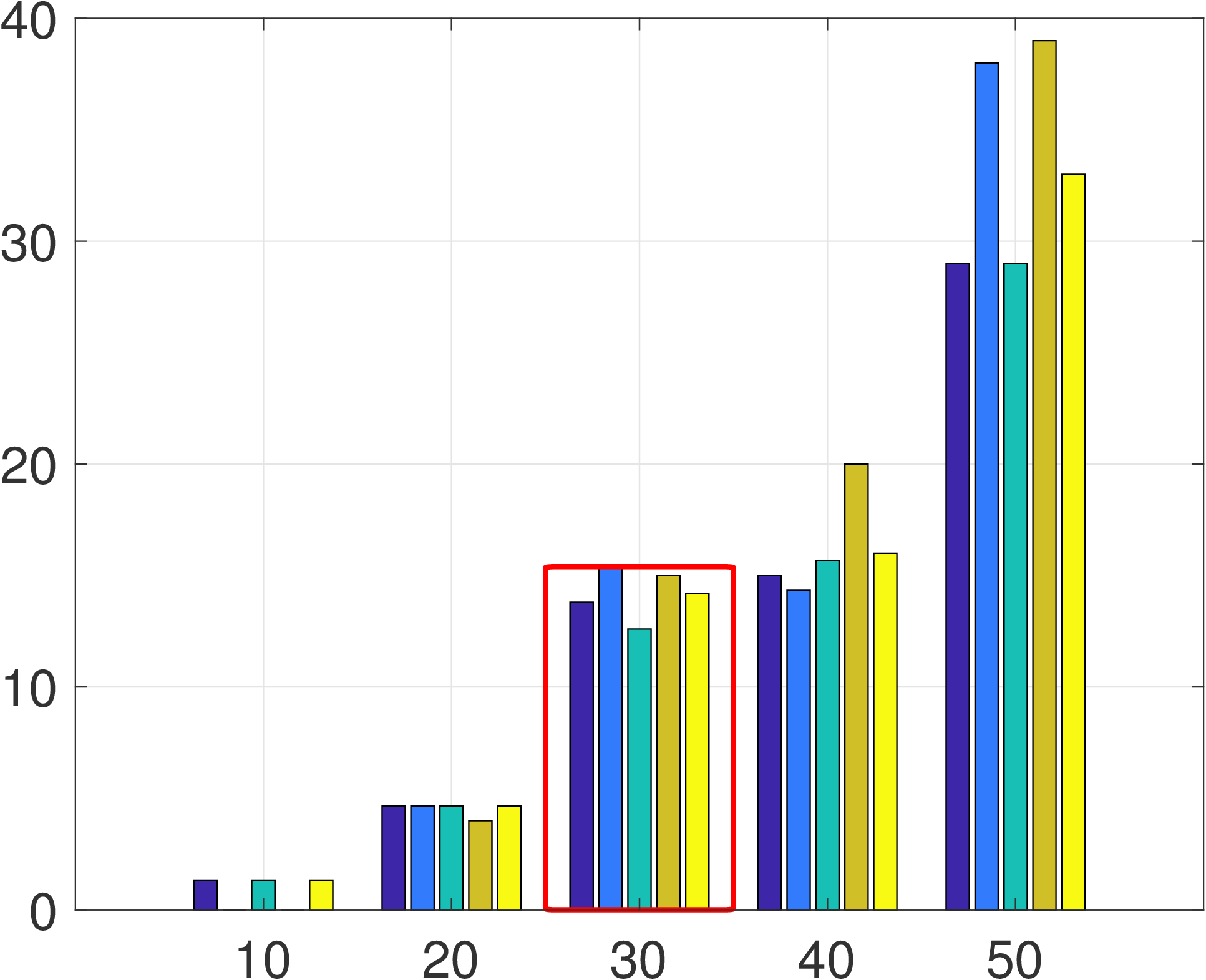}
\caption{bottleneck evacuation}
\end{subfigure}
\begin{subfigure}{0.21\textwidth}
\centering
\includegraphics[width = \textwidth]{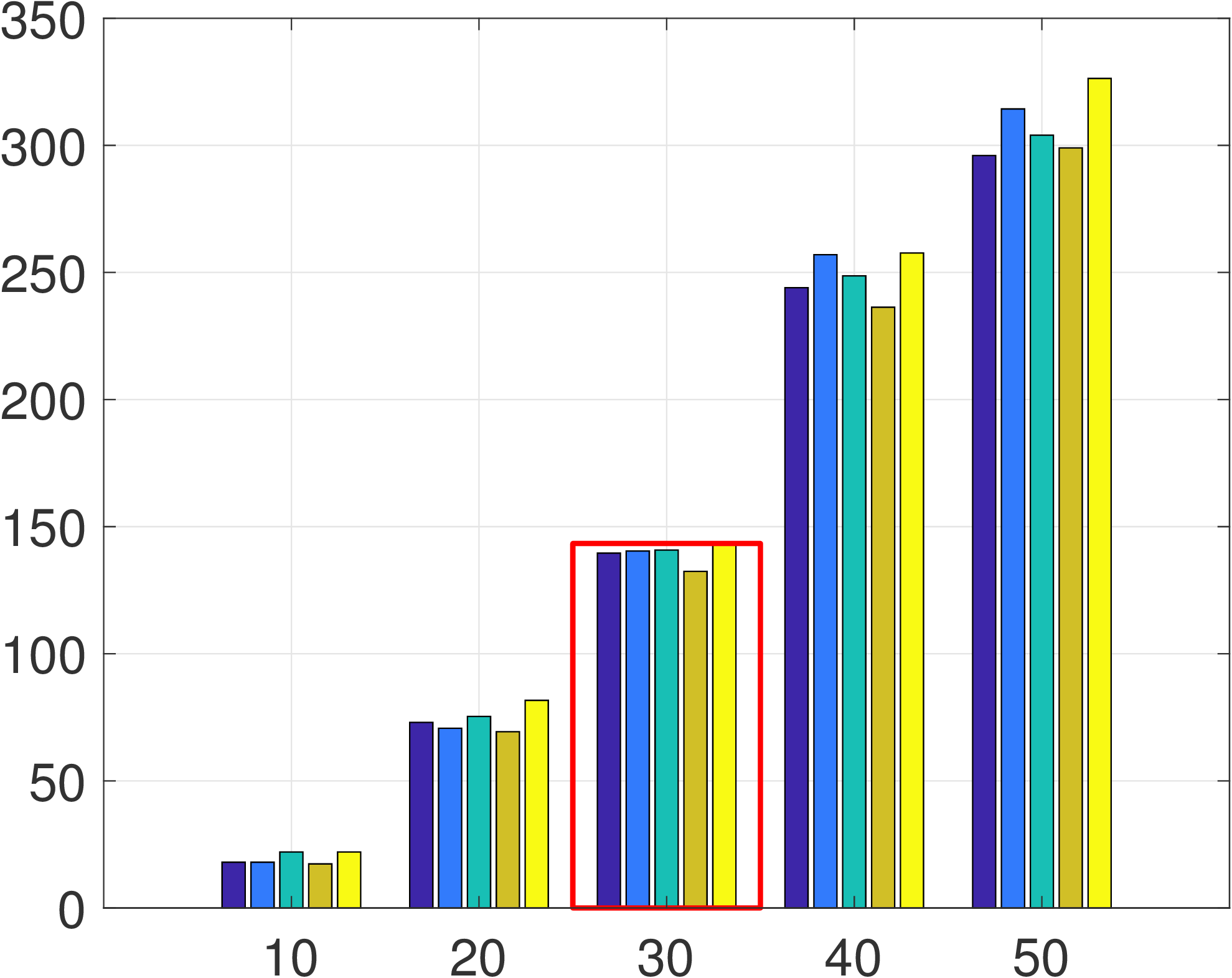}
\caption{bottleneck evacuation 2}
\end{subfigure}

\begin{subfigure}{0.21\textwidth}
\centering
\includegraphics[width = \textwidth]{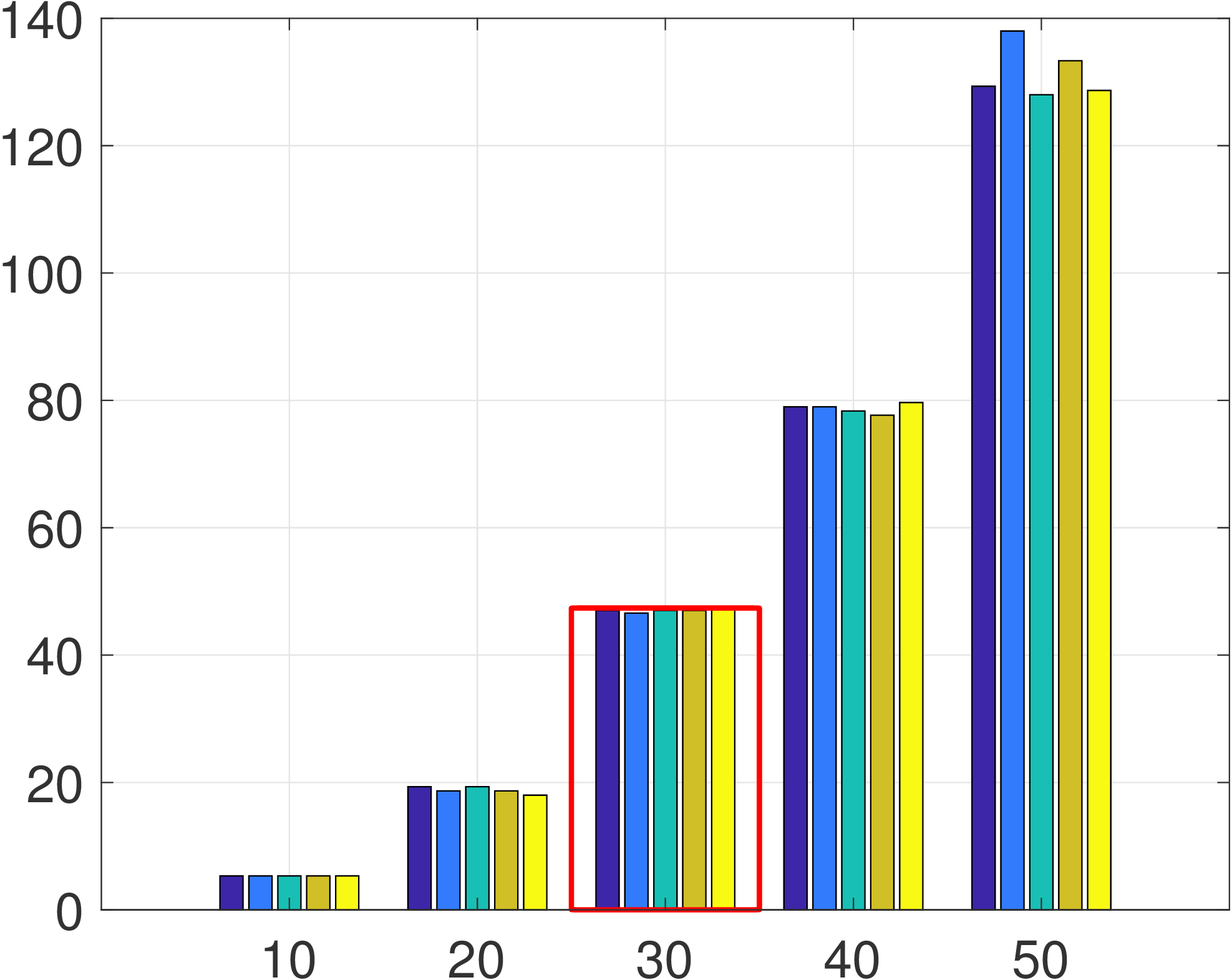}
\caption{bottleneck squeeze}
\end{subfigure}
\begin{subfigure}{0.21\textwidth}
\centering
\includegraphics[width = \textwidth]{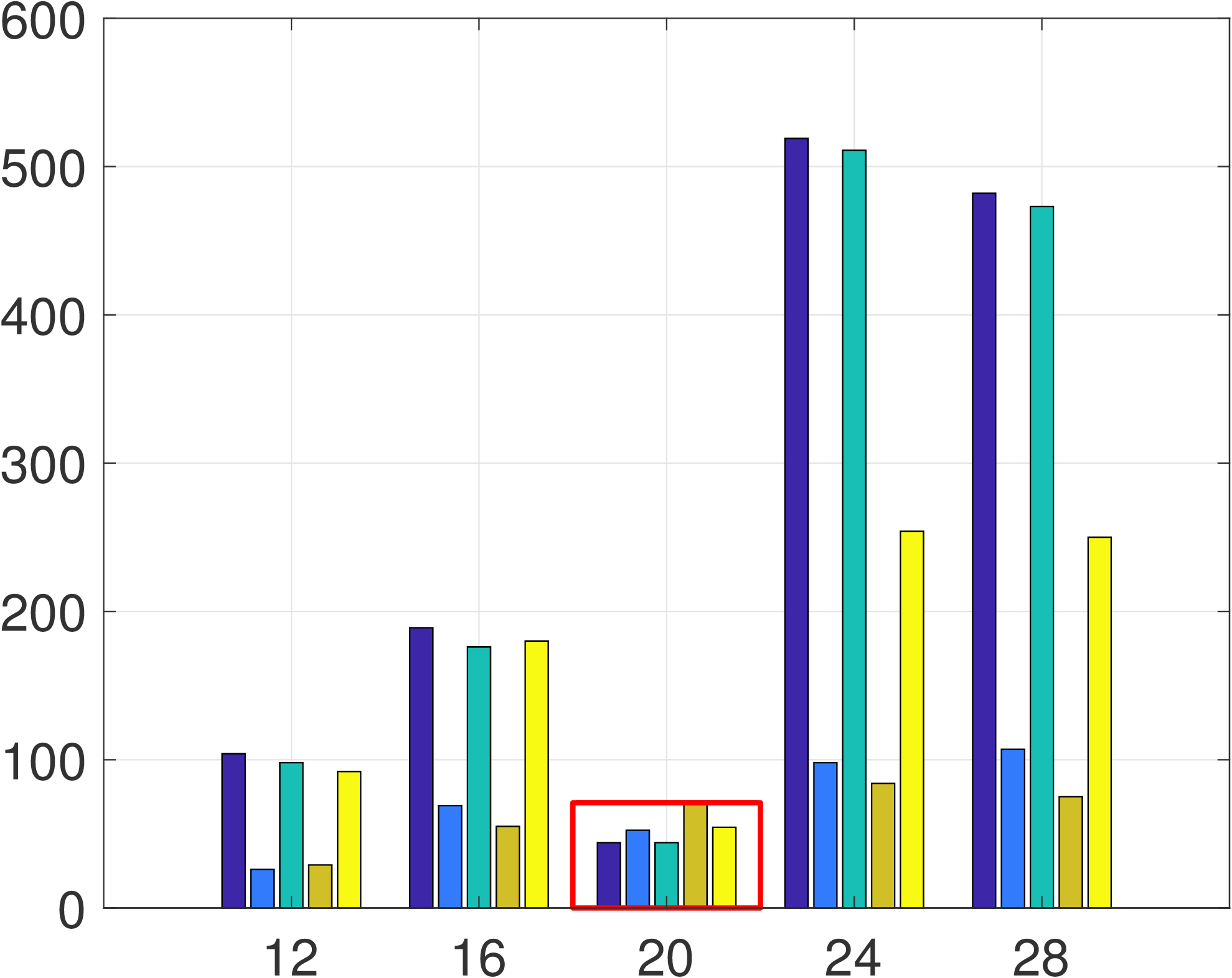}
\caption{concentric circles}
\end{subfigure}

\begin{subfigure}{0.21\textwidth}
\centering
\includegraphics[width = \textwidth]{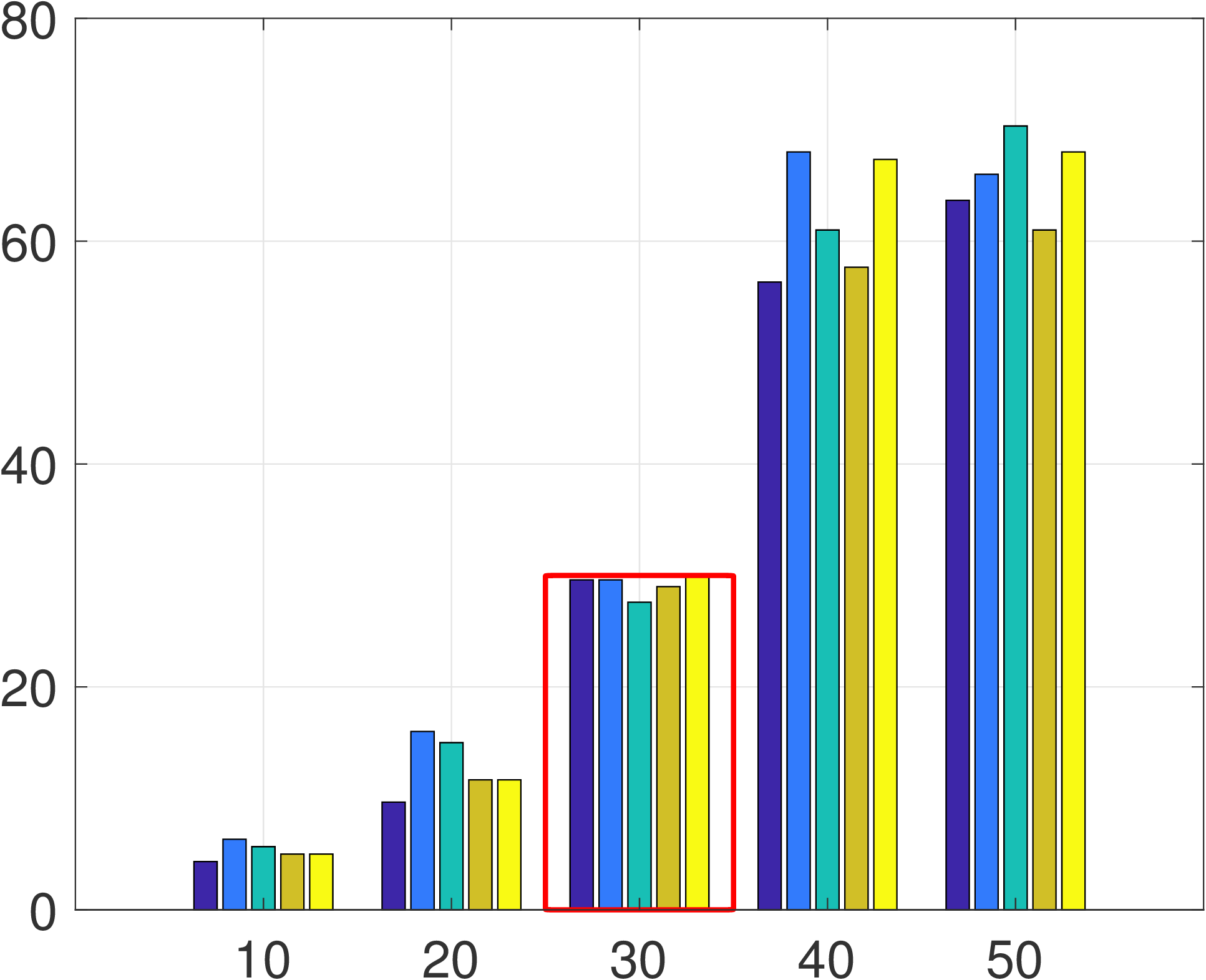}
\caption{hallway-two-way}
\end{subfigure}
\begin{subfigure}{0.21\textwidth}
\centering
\includegraphics[width = \textwidth]{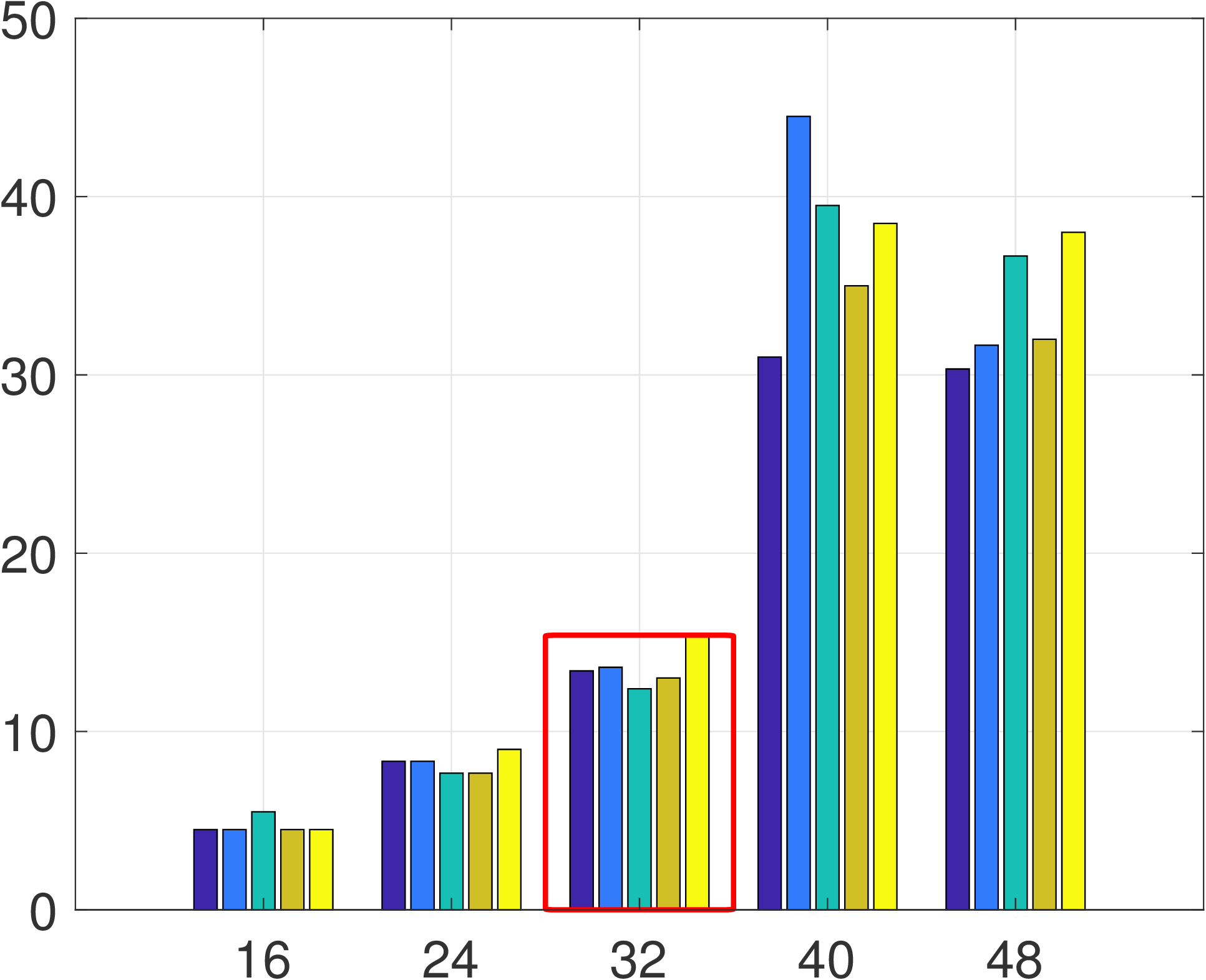}
\caption{hallway-four-way}
\end{subfigure}

\caption{Number of agent-agent collisions over 6 scenarios using UKS optimization for extended evaluation. See Fig.~\ref{fig:DTW_varying_density} for additional caption details.}
\label{fig:AA_varying_density}
\vspace{-0.25in}
\end{figure}

\begin{figure}[t]
\centering

\begin{subfigure}{0.40\textwidth}
\centering
\includegraphics[width = \textwidth,trim={1cm 1.25cm 1cm 2cm},clip]{images/legend.png}
\end{subfigure}

\begin{subfigure}{0.21\textwidth}
\centering
\includegraphics[width = \textwidth]{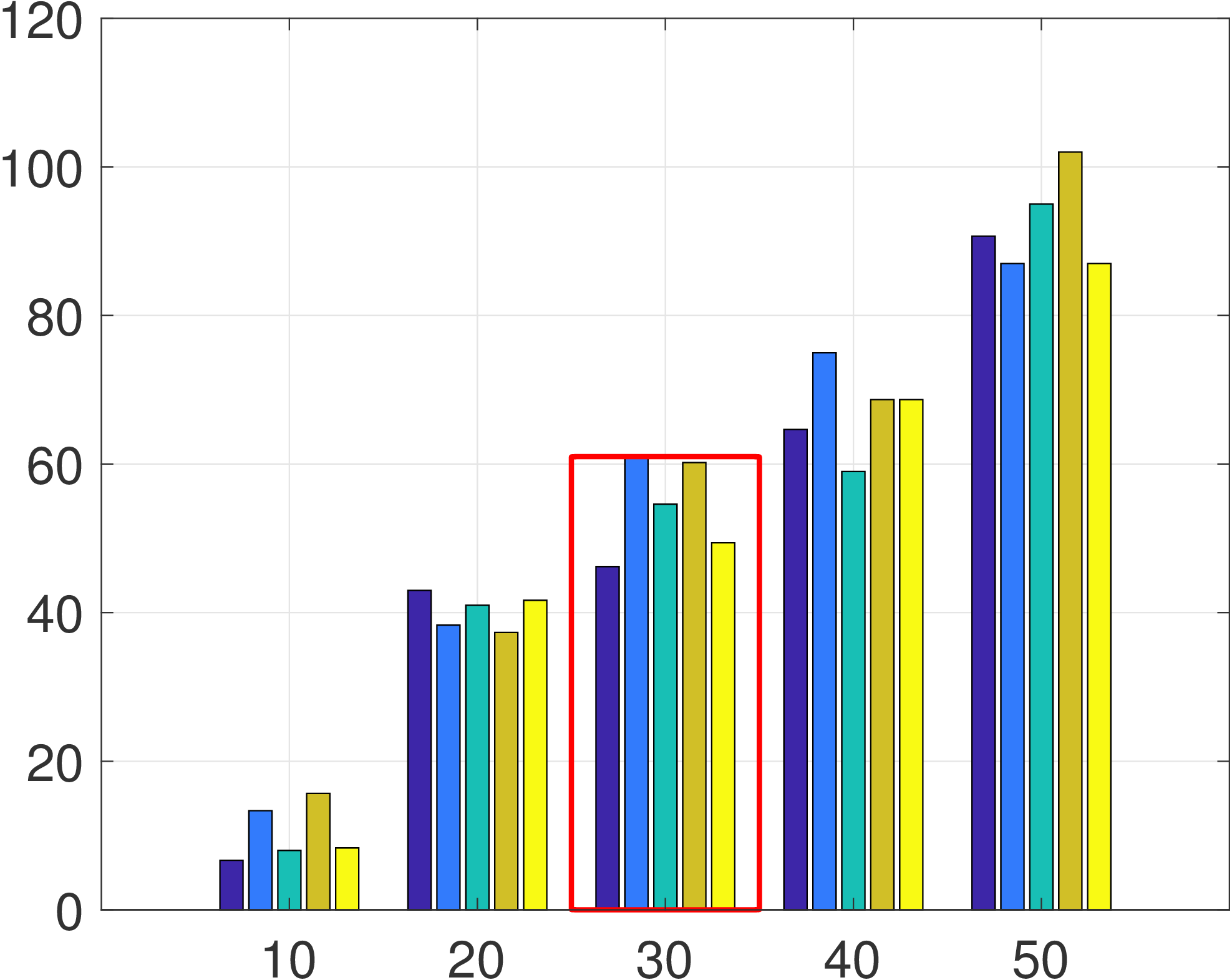}
\caption{bottleneck evacuation}
\end{subfigure}
\begin{subfigure}{0.21\textwidth}
\centering
\includegraphics[width = \textwidth]{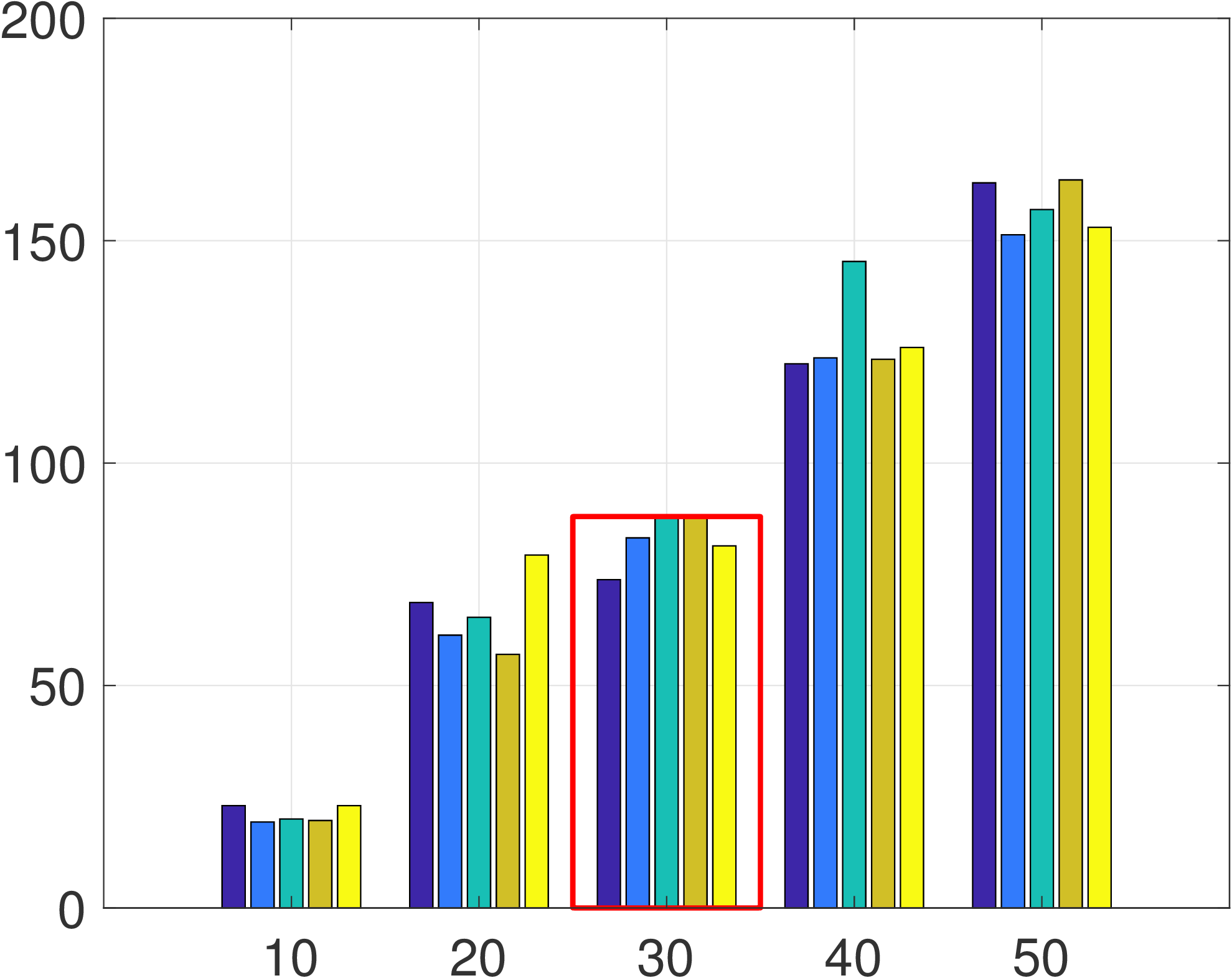}
\caption{bottleneck evacuation 2}
\end{subfigure}

\begin{subfigure}{0.21\textwidth}
\centering
\includegraphics[width = \textwidth]{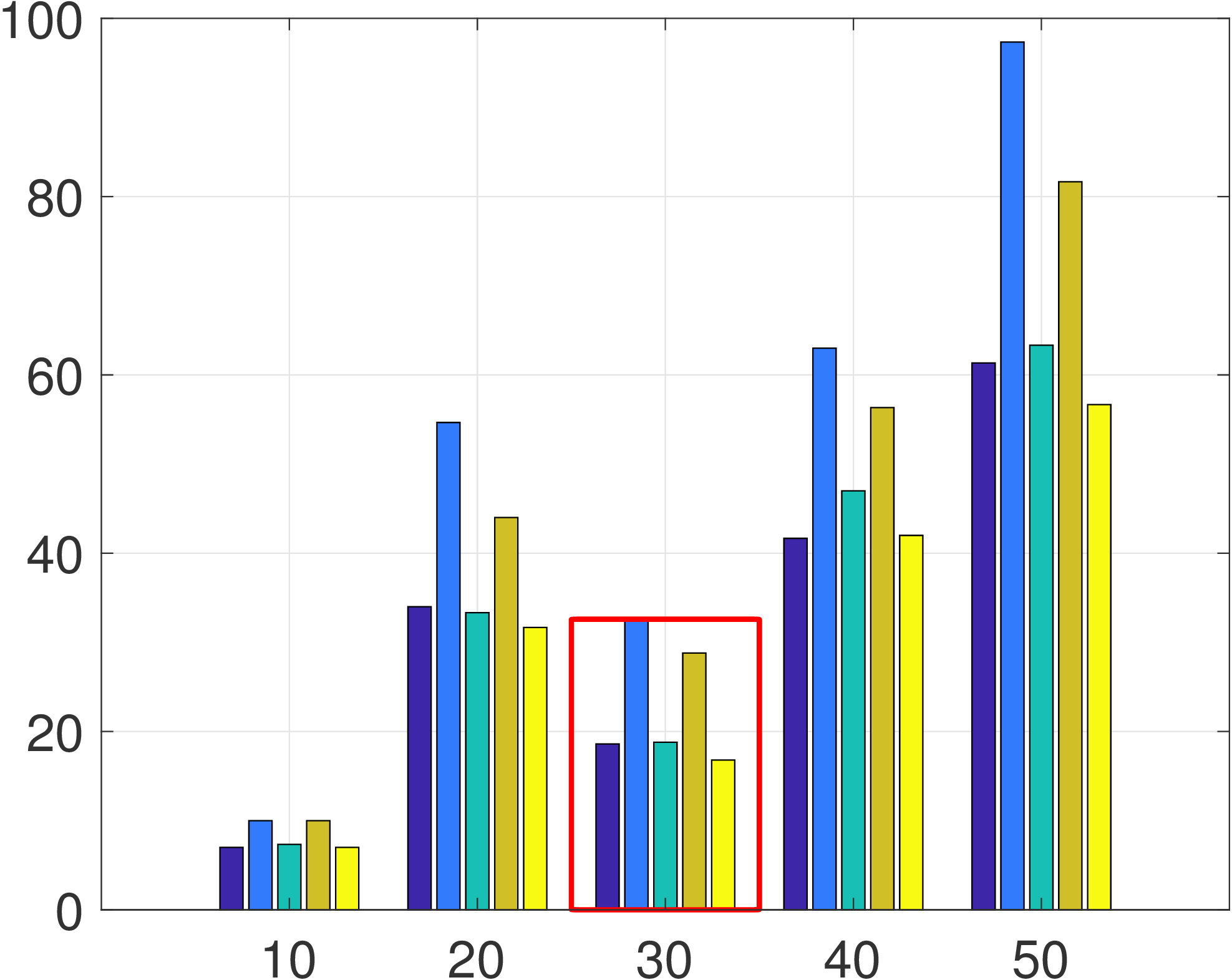}
\caption{bottleneck squeeze}
\end{subfigure}
\begin{subfigure}{0.21\textwidth}
\centering
\includegraphics[width = \textwidth]{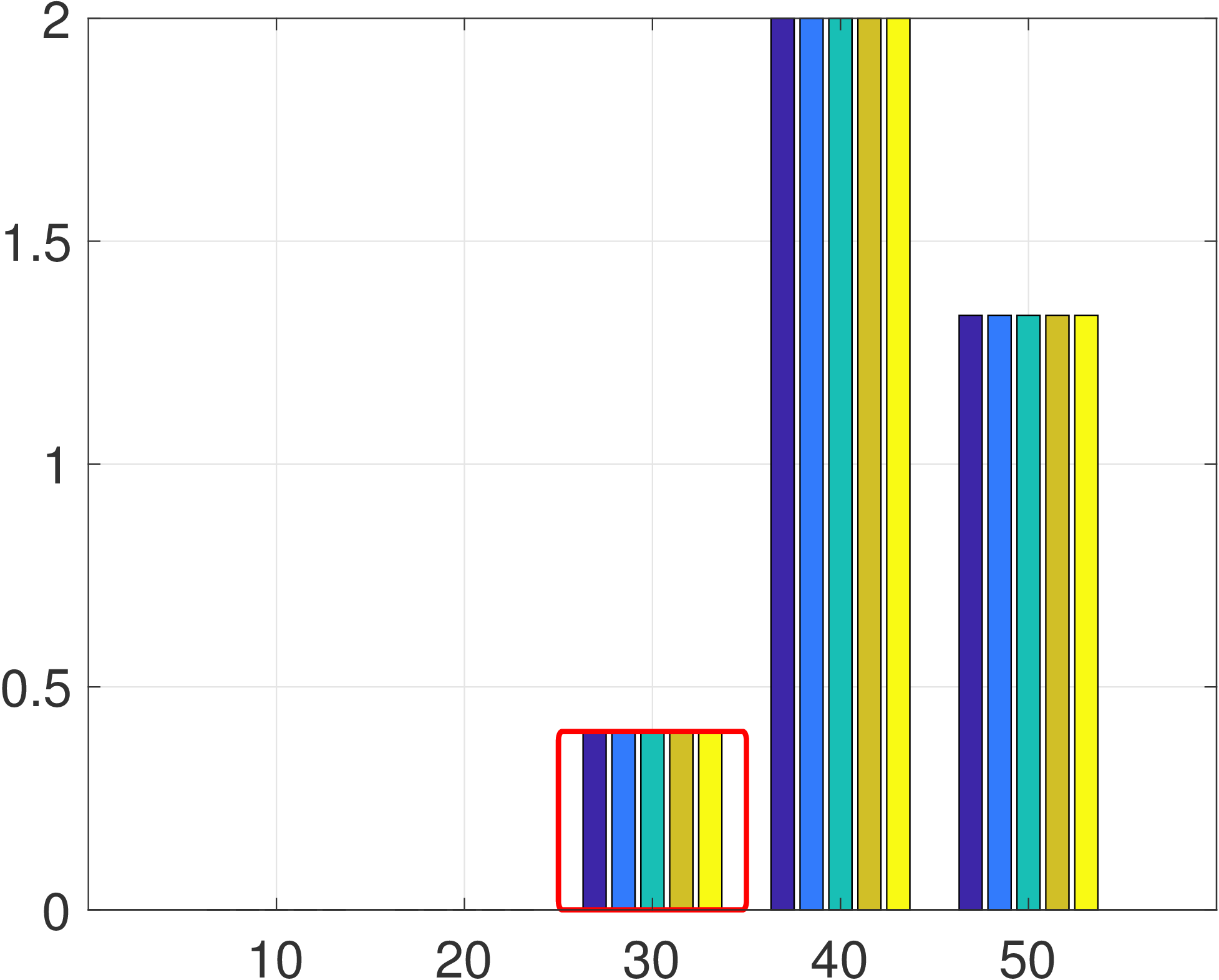}
\caption{hallway-two-way}
\end{subfigure}

\caption{Number of agent-obstacle collisions over 4 scenarios using UKS optimization for extended evaluation. See Fig.~\ref{fig:DTW_varying_density} for additional caption details.}
\label{fig:AO_varying_density}
\vspace{-0.25in}
\end{figure}

\textbf{Experimental Results}. Results of experiments in the first, matching-agent-density setting, are summarized in Table~\ref{Tab:dtw_basic}, ~\ref{Tab:aa_basic}, ~\ref{Tab:ao_basic}. The average rankings of different methods can be used to ascertain relative performance and are presented in Table ~\ref{Tab:average_rank}.   Table~\ref{Tab:cpu_time} shows the computational time of different evaluated approaches.  Finally, results of evaluations across different train-test agent densities are shown in Figures \ref{fig:DTW_varying_density}, \ref{fig:AA_varying_density} and \ref{fig:AO_varying_density}. We only show evaluations for one of the optimization approaches, the UKS, for brevity and because other approaches follow similar trends.

%--------------------------------------------------------------------------

%--------------------------------------------------------

%--------------------------------------------------------

%---------------------------------------------------------------------------------
\begin{table*}
	\small
\centering
\caption{Missing 30\% of frames, Average Relative DTW Distance in Percentage for basic evaluation. Red color indicates the best method in each scenario, followed by the second best in blue. Best viewed in color.}
  \begin{tabular}{l|r|rr|rr|R{1cm}R{1cm}|rr|R{1cm}R{1cm}}
    \toprule
    \multirow{2}{*}{Scenario} & 
      {MPA} &
      \multicolumn{2}{c}{NN } \vline &
      \multicolumn{2}{c}{GP-fed-NN } \vline &
      \multicolumn{2}{c}{LinComb GP+NN} \vline &
      \multicolumn{2}{c}{GP} \vline &
      \multicolumn{2}{c}{GP-fed-ORCA} \\
      & {} & {IPM} & {UKS} &  {IPM} & {UKS} & {IPM} & {UKS} & {IPM} & {UKS} & {IPM} & {UKS}\\
      \midrule
      
    bottleneck-evacuation     & 22.2   & 11.4   & 18.1   & 9.2    & 12.0   & 10.2   & 10.4  & {\color{red}\bf9.0}     & 9.1   & {\color{red}\bf9.0}   & 9.3  \\
    bottleneck-evacuation-2    & 20.9   & 16.3   & 18.5   & 12.7    & 14.4   & 11.4   & 13.1  & {\color{red}\bf9.3}   & {\color{blue}\bf11.1}  & 12.3  & 14.1  \\
    bottleneck-squeeze       & 19.2   & 3.7    & 4.4    & {\color{red}\bf2.3}    & 3.1    & {\color{blue}\bf2.6}    & 2.8   & 3.1   & 3.1   & 3.3   & 3.1   \\
    concentric-circles 	        & 47.2   & {\color{blue}\bf14.8}   & 42.5   & {\color{red}\bf11.2}   & 46.2   & 17.2   & 36.6  & 17.1  & 35.8  & 37.3  & 34.0  \\
    hallway-two-way           & 16.9   & 17.4   & 13.2   & 13.7   & {\color{red}\bf11.2}   & 14.3   & 13.5  & 15.3  & 14.1  & 14.8  & {\color{blue}\bf12.6}  \\
    hallway-four-way          & 18.0   & 25.0   & 17.3   & 15.5   & 14.9   & 15.0   & 14.1  & 12.6  & {\color{red}\bf12.0}    & 12.9  & {\color{blue}\bf12.3}  \\
    \bottomrule
  \end{tabular}
% 	\begin{tabular}{l|S|SS|SS|SS|SS|SS}
%     \toprule
%     \multirow{2}{*}{Scenario} & 
%       {MPA} &
%       \multicolumn{2}{c}{NN } \vline &
%       \multicolumn{2}{c}{GP-fed NN } \vline &
%       \multicolumn{2}{c}{LinComb GP+NN} \vline &
%       \multicolumn{2}{c}{GP} \vline &
%       \multicolumn{2}{c}{ORCA driven by GP} \\
%       & {} & {IPM} & {UKS} &  {IPM} & {UKS} & {IPM} & {UKS} & {IPM} & {UKS} & {IPM} & {UKS}\\
%       \midrule
    
%     bottleneck-evacuation ($200 \times 160$)    & 22.2   & 11.41   & 18.17   & 9.22    & 12.08   & 10.25   & 10.43  & {\color{red}\bf 9}     & 9.1   & {\color{red}\bf 9}   & 9.35  \\
%     bottleneck-evacuation-2  ($100 \times 80$)  & 20.9   & 16.31   & 18.58   & 12.7    & 14.46   & 11.45   & 13.15  & {\color{red}\bf 9.3}   & {\color{blue}\bf 11.1}  & 12.38  & 14.13  \\
%     bottleneck-squeeze ($200 \times 200$)       & 19.2   & 3.79    & 4.49    & {\color{red}\bf 2.32}    & 3.16    & {\color{blue}\bf 2.63}    & 2.87   & 3.1   & 3.1   & 3.34   & 3.19   \\
%     concentric-circles ($20 \times 20$)	        & 47.2   & {\color{blue}\bf 14.88}   & 42.55   & {\color{red}\bf 11.28}   & 46.22   & 17.21   & 36.67  & 17.1  & 35.8  & 37.33  & 34.04  \\
%     hallway-two-way ($200 \times 200$)          & 16.9   & 17.41   & 13.25   & 13.76   & {\color{red}\bf 11.21}   & 14.35   & 13.53  & 15.3  & 14.1  & 14.88  & {\color{blue}\bf 12.68}  \\
%     hallway-four-way ($200 \times 200$)         & 18.0   & 25.08   & 17.34   & 15.52   & 14.94   & 15.07   & 14.15  & 12.6  & {\color{red}\bf 12}    & 12.99  & {\color{blue}\bf 12.32}  \\
%     \bottomrule
%   \end{tabular}
  \label{Tab:dtw_basic}
  \vspace{-0.25in}
\end{table*}

%-----------------------------------------------------------------------------------------------------
\begin{table*}
	\small
\centering
\caption{Missing 30\% of frames, Agent-Agent Collisions for basic evaluation. Red color indicates the best method in each scenario, followed by the second best in blue. Best viewed in color.}
  \begin{tabular}{l|r|rr|rr|R{1cm}R{1cm}|rr|R{1cm}R{1cm}}
    \toprule
    \multirow{2}{*}{Scenario} & 
      {MPA} &
      \multicolumn{2}{c}{NN} \vline &
      \multicolumn{2}{c}{GP-fed-NN} \vline &
      \multicolumn{2}{c}{LinComb GP+NN} \vline &
      \multicolumn{2}{c}{GP} \vline &
      \multicolumn{2}{c}{GP-fed-ORCA} \\
      & {} & {IPM} & {UKS} &  {IPM} & {UKS} & {IPM} & {UKS} & {IPM} & {UKS} & {IPM} & {UKS}\\
      \midrule
    
    bottleneck-evacuation     & {\color{red}\bf 10.6}	& 16.2	 & 15.4	  & 17.2   & 15.0	    & 14.6	  & {\color{blue}\bf 12.6}	& 14.2	 & 13.8	  & 16.0	   & 14.2  \\
    bottleneck-evacuation-2   & 147.0	& 159.8	 & 140.4  & 157.0	 & {\color{red}\bf 132.4}	& {\color{blue}\bf 139.4}	  & 140.8	& 144.8	 & 139.6  & 158.6  & 143.4 \\
    bottleneck-squeeze       & 132.0	& 49.0	 & {\color{red}\bf 46.6}	  & 47.4   & 47.0	    & {\color{red}\bf 46.6}	  & 47.0	    & 47.0     & 47.0	  & 47.4   & 47.4  \\
    concentric-circles        & 63.4	& {\color{red}\bf 18.4}	 & 52.4	  & {\color{blue}\bf 19.6}   & 71.0	    & 26.0	  & 44.0	    & 26.0	 & 44.0	  & 67.0	   & 54.4  \\
    hallway-two-way          & 53.6	& 30.4	 & 29.6	  & {\color{blue}\bf 27.8}   & 29.0	    & 29.8	  & {\color{red}\bf 27.6}		& 29.8	 & 29.6	  & 29.0	   & 30.0    \\
    hallway-four-way        & 35.8	& 13.6	 & 13.6	  & {\color{blue}\bf 12.2}  & 13.0	    & {\color{red}\bf 10.6}	  & 12.4	& 13.6	 & 13.4	  & 15.0	   & 15.4  \\
    \bottomrule
  \end{tabular}
%   \begin{tabular}{l|S|SS|SS|SS|SS|SS}
%     \toprule
%     \multirow{2}{*}{Scenario (environment width x height)} & 
%       {MPA} &
%       \multicolumn{2}{c}{NN} \vline &
%       \multicolumn{2}{c}{GP-fed NN} \vline &
%       \multicolumn{2}{c}{LinComb GP+NN} \vline &
%       \multicolumn{2}{c}{GP} \vline &
%       \multicolumn{2}{c}{ORCA driven by GP} \\
%       & {} & {IPM} & {UKS} &  {IPM} & {UKS} & {IPM} & {UKS} & {IPM} & {UKS} & {IPM} & {UKS}\\
%       \midrule
    
%     bottleneck-evacuation ($200 \times 160$)    & {\color{red}\bf 10.6}	& 16.2	 & 15.4	  & 17.2   & 15	    & 14.6	  & {\color{blue}\bf 12.6}	& 14.2	 & 13.8	  & 16	   & 14.2  \\
%     bottleneck-evacuation-2  ($100 \times 80$)  & 147	& 159.8	 & 140.4  & 157	 & {\color{red}\bf 132.4}	& {\color{blue}\bf 139.4}	  & 140.8	& 144.8	 & 139.6  & 158.6  & 143.4 \\
%     bottleneck-squeeze ($200 \times 200$)       & 132	& 49	 & {\color{red}\bf 46.6}	  & 47.4   & 47	    & {\color{red}\bf 46.6}	  & 47	    & 47     & 47	  & 47.4   & 47.4  \\
%     concentric-circles ($20 \times 20$)	        & 63.4	& {\color{red}\bf 18.4}	 & 52.4	  & {\color{blue}\bf 19.6}   & 71	    & 26	  & 44	    & 26	 & 44	  & 67	   & 54.4  \\
%     hallway-two-way ($200 \times 200$)          & 53.6	& 30.4	 & 29.6	  & {\color{blue}\bf 27.8}   & 29	    & 29.8	  & {\color{red}\bf 27.6}	& 29.8	 & 29.6	  & 29	   & 30    \\
%     hallway-four-way ($200 \times 200$)         & 35.8	& 13.6	 & 13.6	  & {\color{blue}\bf 12.2 }  & 13	    & {\color{red}\bf 10.6}	  & 12.4	& 13.6	 & 13.4	  & 15	   & 15.4  \\
%     \bottomrule
%   \end{tabular}
  \label{Tab:aa_basic}
\end{table*}

%---------------------------------------------------------------------------------------------------------
\begin{table*}
	\small
\centering
\caption{Missing 30\% of frames, Average-Obstacle Collisions for basic evaluation. Red color indicates the best method in each scenario, followed by the second best in blue. Best viewed in color.}
  \begin{tabular}{l|r|rr|rr|R{1cm}R{1cm}|rr|R{1cm}R{1cm}}
    \toprule
    \multirow{2}{*}{Scenario} & 
      {MPA} &
      \multicolumn{2}{c}{NN} \vline &
      \multicolumn{2}{c}{GP-fed-NN} \vline &
      \multicolumn{2}{c}{LinComb GP+NN} \vline &
      \multicolumn{2}{c}{GP} \vline &
      \multicolumn{2}{c}{GP-fed-ORCA} \\
      & {} & {IPM} & {UKS} &  {IPM} & {UKS} & {IPM} & {UKS} & {IPM} & {UKS} & {IPM} & {UKS}\\
      \midrule
    bottleneck-evacuation       & {\color{blue}\bf 39.2}	 & 52.4	  & 61	  & 45.4   & 60.2  & 46.6	& 54.6	& {\color{red}\bf 37.2}	& 46.2	& 39.4	& 49.4 \\
    bottleneck-evacuation-2      & {\color{red}\bf 46.2}	 & 75.4	  & 83.2  & 76.4   & 88.0	   & 88.8	& 87.6	& {\color{blue}\bf 70.4}	& 73.8	& 89.4	& 81.4 \\
    bottleneck-squeeze          & 28.0	     & 30.8	  & 32.6  & {\color{blue}\bf 7.8}	   & 28.8  & 8.2	& 18.8	& {\color{red}\bf 6.6}	& 18.6	& 13.2	& 16.8 \\
    hallway-two-way             & 0.0	     & 0.4	  & 0.4	  & 0.4	   & 0.4   & 0.4	& 0.4	& 0.4	& 0.4	& 0.4	& 0.4  \\
    hallway-four-way            & 0.0	     & 0.0	  & 0.0	  & 0.0	   & 0.0	   & 0.0	    & 0.0	    & 1.8	& 0.0	    & 3.2	& 0.4  \\
    \bottomrule
  \end{tabular}
%   \begin{tabular}{l|S|SS|SS|SS|SS|SS}
%     \toprule
%     \multirow{2}{*}{Scenario (environment width x height)} & 
%       {MPA} &
%       \multicolumn{2}{c}{NN} \vline &
%       \multicolumn{2}{c}{GP-fed NN} \vline &
%       \multicolumn{2}{c}{LinComb GP+NN} \vline &
%       \multicolumn{2}{c}{GP} \vline &
%       \multicolumn{2}{c}{ORCA driven by GP} \\
%       & {} & {IPM} & {UKS} &  {IPM} & {UKS} & {IPM} & {UKS} & {IPM} & {UKS} & {IPM} & {UKS}\\
%       \midrule
%     bottleneck-evacuation ($200 \times 160$)      & {\color{blue}\bf 39.2}	 & 52.4	  & 61	  & 45.4   & 60.2  & 46.6	& 54.6	& {\color{red}\bf 37.2}	& 46.2	& 39.4	& 49.4 \\
%     bottleneck-evacuation-2  ($100 \times 80$)    & {\color{red}\bf 46.2}	 & 75.4	  & 83.2  & 76.4   & 88	   & 88.8	& 87.6	& {\color{blue}\bf 70.4}	& 73.8	& 89.4	& 81.4 \\
%     bottleneck-squeeze ($200 \times 200$)         & 28	     & 30.8	  & 32.6  & {\color{blue}\bf 7.8}	   & 28.8  & 8.2	& 18.8	& {\color{red}\bf 6.6}	& 18.6	& 13.2	& 16.8 \\
%     hallway-two-way ($200 \times 200$)            & 0	     & 0.4	  & 0.4	  & 0.4	   & 0.4   & 0.4	& 0.4	& 0.4	& 0.4	& 0.4	& 0.4  \\
%     hallway-four-way ($200 \times 200$)           & 0	     & 0	  & 0	  & 0	   & 0	   & 0	    & 0	    & 1.8	& 0	    & 3.2	& 0.4  \\
%     \bottomrule
%   \end{tabular}
  \label{Tab:ao_basic}
\end{table*}

%-----------------------------------------------------------------
\begin{table*}
	\small
\centering
\caption{Average ranks of different combinations over scenarios for basic evaluation. Lower rank is better. Red color indicates the best method for each evaluation approach, followed by the second best in blue. Best viewed in color.}
  \begin{tabular}{l|r|rr|rr|R{1cm}R{1cm}|rr|R{1cm}R{1cm}}
    \toprule
    \multirow{2}{*}{Evaluation} & 
      {MPA} &
      \multicolumn{2}{c}{NN} \vline &
      \multicolumn{2}{c}{GP-fed-NN} \vline &
      \multicolumn{2}{c}{LinComb GP+NN} \vline &
      \multicolumn{2}{c}{GP} \vline &
      \multicolumn{2}{c}{GP-fed-ORCA} \\
      & {} & {IPM} & {UKS} &  {IPM} & {UKS} & {IPM} & {UKS} & {IPM} & {UKS} & {IPM} & {UKS}\\
      \midrule
    DTW            & 10.67   & 8.33   & 8.50   & 4.00     & 6.67  & 4.83  & 5.33   & {\color{red}\bf 3.67}  & {\color{blue}\bf 3.75}  & 5.58  & 4.67 \\
    Agent-Agent    & 8.50     & 8.17   & 5.50   & 5.67  & 5.17  & {\color{blue}\bf 3.58}  & {\color{red}\bf 3.50}    & 5.67  & 4.41  & 8.25  & 7.58 \\
    Agent-Obstacle & {\color{red}\bf 3.30}     & 6.60    & 8.00     & 4.40   & 7.80   & 6.00     & 7.00      & {\color{blue}\bf 4.10}   & 5.00     & 7.10   & 6.70 \\
    \bottomrule
  \end{tabular}
%   \begin{tabular}{l|S|SS|SS|SS|SS|SS}
%     \toprule
%     \multirow{2}{*}{Evaluation} & 
%       {MPA} &
%       \multicolumn{2}{c}{NN} \vline &
%       \multicolumn{2}{c}{GP-fed NN} \vline &
%       \multicolumn{2}{c}{LinComb GP+NN} \vline &
%       \multicolumn{2}{c}{GP} \vline &
%       \multicolumn{2}{c}{ORCA driven by GP} \\
%       & {} & {IPM} & {UKS} &  {IPM} & {UKS} & {IPM} & {UKS} & {IPM} & {UKS} & {IPM} & {UKS}\\
%       \midrule
%     DTW            & 10.67   & 8.33   & 8.5   & 4     & 6.67  & 4.83  & 5.33   & {\color{red}\bf 3.67}  & {\color{blue}\bf 3.75}  & 5.58  & 4.67 \\
%     Agent-Agent    & 8.5     & 8.17   & 5.5   & 5.67  & 5.17  & {\color{blue}\bf 3.58}  & {\color{red}\bf 3.5}    & 5.67  & 4.41  & 8.25  & 7.58 \\
%     Agent-Obstacle & {\color{red}\bf 3.3}     & 6.6    & 8     & 4.4   & 7.8   & 6     & 7      & {\color{blue}\bf 4.1}   & 5     & 7.1   & 6.7 \\
%     \bottomrule
%   \end{tabular}
  \label{Tab:average_rank}
\end{table*}

%--------------------------------------------------------------------------------
\begin{table*}
	\small
\centering
\caption{Computational time, in seconds, for basic evaluation. Note that in this table, the time of IPM and UKS is sequentially accumulated over trajectories, while in practice, trajectories could be optimized in parallel when using IPM and UKS.}
  \begin{tabular}{l|r|rr|rr|R{1cm}R{1cm}|rr|R{1cm}R{1cm}}
    \toprule
    \multirow{2}{*}{Scenario} & {MPA} &
      \multicolumn{2}{c}{NN } \vline &
      \multicolumn{2}{c}{GP-fed-NN } \vline &
      \multicolumn{2}{c}{LinComb GP+NN} \vline &
      \multicolumn{2}{c}{GP} \vline &
      \multicolumn{2}{c}{GP-fed-ORCA} \\
      & {} & {IPM} & {UKS} &  {IPM} & {UKS} & {IPM} & {UKS} & {IPM} & {UKS} & {IPM} & {UKS}\\
      \midrule
    
    bottleneck-evacuation   &1343.6	&887.9	&304.4	&887.5	&352.7	&922.2	&337.9	&881.1	&353.4	&681.1	&144.4 \\
    bottleneck-evacuation-2 &521.1	&290.1	&201.4	&321.2	&233.8	&308.7	&222.4	&320.0	&234.7	&174.7	&87.7  \\
    bottleneck-squeeze      &738.8	&746.8	&283.9	&769.3	&329.6	&637.5	&313.4	&615.7	&331.2	&475.5	&136.4  \\
    concentric-circles	    &29.5	&173.0	&143.8	&191.6	&162.0	&178.0	&148.5	&192.1	&162.2	&85.2	&56.3  \\ 
    hallway-two-way         &876.3	&841.5	&293.9	&876.4	&342.6	&885.8	&324.9	&880.7	&343.7	&694.7	&143.2 \\ 
    hallway-four-way        &1135.9	&901.9	&319.1	&941.4	&369.5	&944.4	&350.6	&944.4	&368.8	&729.7	&147.2 \\
    
    \bottomrule
  \end{tabular}
%   \begin{tabular}{l|S|SS|SS|SS|SS|SS}
%     \toprule
%     \multirow{2}{*}{Scenario (subsampled frames)} & {MPA} &
%       \multicolumn{2}{c}{NN } \vline &
%       \multicolumn{2}{c}{GP-fed NN } \vline &
%       \multicolumn{2}{c}{LinComb GP+NN} \vline &
%       \multicolumn{2}{c}{GP} \vline &
%       \multicolumn{2}{c}{ORCA driven by GP} \\
%       & {} & {IPM} & {UKS} &  {IPM} & {UKS} & {IPM} & {UKS} & {IPM} & {UKS} & {IPM} & {UKS}\\
%       \midrule
    
%     bottleneck-evacuation ($101$)   &1343.6	&887.94	&304.45	&887.51	&352.72	&922.26	&337.94	&881.14	&353.45	&681.13	&144.47 \\
%     bottleneck-evacuation-2 ($53$)  &521.17	&290.12	&201.48	&321.2	&233.87	&308.75	&222.48	&320.03	&234.73	&174.73	&87.79  \\
%     bottleneck-squeeze ($94$)       &738.8	&746.89	&283.97	&769.3	&329.68	&637.58	&313.49	&615.78	&331.21	&475.56	&136.4  \\
%     concentric-circles ($37$)	    &29.57	&173.08	&143.87	&191.66	&162.01	&178.02	&148.58	&192.17	&162.27	&85.29	&56.37  \\ 
%     hallway-two-way ($101$)         &876.32	&841.54	&293.91	&876.47	&342.6	&885.86	&324.96	&880.77	&343.78	&694.78	&143.29 \\ 
%     hallway-four-way($101$)         &1135.9	&901.93	&319.15	&941.43	&369.5	&944.4	&350.63	&944.44	&368.88	&729.72	&147.26 \\
    
%     \bottomrule
%   \end{tabular}
  \label{Tab:cpu_time}
\end{table*}

%--------------------------------------------------------
\section{Discussion}
 
For the basic evaluations where the agent density of the test set matches that of the training set, we can see from Table \ref{Tab:dtw_basic}, \ref{Tab:aa_basic} and \ref{Tab:ao_basic} that (i) for DTW score, GP-fed-NN, GP and GP-fed-ORCA perform better than other priors. This illustrates that GP is important for regulating the trajectory to roughly follow the movement (flow) pattern. (ii) In terms of the number of agent-agent collisions, GP-fed-NN, LinComb GP+NN perform best, followed by NN. This demonstrates that pure GP alone is not sufficient to guarantee the desired collision avoidance, and has to be augmented by another policy learner like NN to achieve the least number of agent-agent collisions. Surprisingly, for this metric GP-fed-ORCA does not give satisfactory results, which might stem from our complex training/testing datasets, containing densely packed agents (especially for bottleneck-evacuation-2 scenario) such that the permitted set of ORCA is prone to be empty under such agent density. (iii) For the number of agent-obstacle collisions, MPA and GP perform better than others, indicating the importance of the global flow patterns in order to avoid stationary environmental obstacles.

Tab.~\ref{Tab:average_rank} suggests that, as the optimization approach, IPM is slightly better than UKS and MPA for DTW; on the other hand, UKS achieves the fewest agent-agent collisions, lower than IPM and MPA.  MPA yields the fewer agent-obstacle collisions than IPM and UKS. However, in terms of computational complexity, Tab.~\ref{Tab:cpu_time} shows that MPA is the most expensive approach since it includes pair-wise constraint. In addition, UKS is cheaper than IPM, in that UKS exploits fast forward-backward computations specific to the trajectory domain. The lack of conformability in the above observations implies that there might not exist a dominant optimization algorithm and one might need to trade off when choosing the optimizations.
 
Evaluation across different agent densities, shown in Fig \ref{fig:DTW_varying_density}, \ref{fig:AA_varying_density} and \ref{fig:AO_varying_density}, provides an insight into the important generalization ability and robustness of proposed approaches.  We can see that (i) for DTW, GP-fed-ORCA performs best except for concentric-circle scenario, where NN presents least DTW distance with respect to the ground truth test trajectories. The reason might be that in concentric-circle scenario, agents are symmetrically placed along a circle, leading to observation patterns similar across all agents and, hence,  local NN policy patterns that can be learned by reusing (sharing) the data across agents. On the other hand, global GP fails to reuse the data, leading to a weaker generalization model.  (ii) There is a general increasing monotonic trend in the number of collisions across densities. This is an intuitive outcome, however without a clear winner in terms of collisions except for concentric-circle scenario. For this scenario, again, the ability to reuse data across agents, due to symmetry, may lead to better generalization of the local NN approach. (iii) Models in the matching density settings outperform those in the mismatched density settings in terms of the agent-agent collisions only for the concentric-circle scenario, while in terms of the agent-obstacle collisions this happens for the bottleneck-squeeze scenario. These results suggest that varying density impact the performance of different models in a reasonable and predictable manner.
 
 Overall, the above observations emphasize the importance of the global flow priors, embodied in the GP model, in our multi-agent trajectory optimization framework. Local collision avoidance priors, manifested through the NN model, have lesser than expected yet still measurable impact.  We plan to further investigate these factors as well as the generalization across environments and considerations of exploiting advanced trajectory learning algorithms where the dynamics of the trajectory and the cost of making an inference could be implicitly learned, in our future work.

%%%%%%%%%% Merge with supplemental materials %%%%%%%%%%
\pagebreak
\begin{center}
\textbf{\large Supplemental Materials}
\end{center}
%%%%%%%%%% Merge with supplemental materials %%%%%%%%%%
%%%%%%%%%% Prefix a "S" to all equations, figures, tables and reset the counter %%%%%%%%%%
\setcounter{equation}{0}
\setcounter{figure}{0}
\setcounter{table}{0}
\setcounter{page}{1}
\makeatletter
\renewcommand{\theequation}{S\arabic{equation}}
\renewcommand{\thefigure}{S\arabic{figure}}
\renewcommand{\thetable}{S\arabic{table}}
%%%%%%%%%% Prefix a "S" to all equations, figures, tables and reset the counter %%%%%%%%%%

In this supplementary material, we describe details on the optimization methods we used in the main manuscript and provide some additional details on the experiments. We use the same notation that was used in the main manuscript.

%--------------------------------------------------------
\section{Global Objective}
We define our global objective function as a combination of local and global priors we described in the main manuscript. First, we summarize the local and global prior terms we presented in the main manuscript. 
\begin{align}
E_{gt}^{i}( \mathbf{x}^{i} | \mathbf{o}^{i} ) &= \sum_{t} u_{t}^{i} \| \mathbf{x}_{t}^{i} - \mathbf{o}_{t}^{i} \|^{2} \\
E_{kn}^{i}( \mathbf{x}^{i} ) &= C_{kn} \sum_{t} \| \mathbf{x}_{t}^{i} - \mathbf{x}_{t-1}^{i} \|^{2} \\
E_{mv}^{i}( \mathbf{x}^{i} ) &=
\begin{cases}
0 &\, \text{if } \| \mathbf{v}_{t}^{i} \| \le C_{mv} \\
%0 &\, \text{if } \| \mathbf{x}_{t}^{i} - \mathbf{x}_{t-1}^{i} \| \le C_{mv} \\
\infty &\, \text{otherwise}
\end{cases}, \forall t = 1..T \\%\label{eq:max_speed_contraint}.
E_{dg}^{i}( \mathbf{x}^{i} | \mathbf{o}^{i} ) 
	&= \lambda \sum_{t} \left\Vert \mathbf{x}_{t}^{i} - \mathbf{x}_{t-1}^{i} \right. \nonumber \\
    &\quad \left. - \Delta t \cdot f_{NN}\left(\mathbf{o}_{NN}^{i},\mu_{GP}\left(\mathbf{o}_{t}^{i},\theta_{GP}\right) \right) \right\Vert^{2}
\label{eq:S_new_unary_nonlin_loc}
\end{align}
where $\lambda = 1 / (\sigma_{NN}^2 \Delta t^{2})$.
Combining all unary prior terms, we obtain the final global objective:
\begin{align}
\mathbf{\hat{X}} 
	%&= \argmin_{\mathbf{X}}\sum_{i} E_{u}^{i}(\mathbf{x}^{i},\mathbf{v}^{i}|\mathbf{o}^{i}), \\
	&= \argmin_{\mathbf{X}}\sum_{i} \left\{ E_{gt}^{i}( \mathbf{x}^{i} | \mathbf{o}^{i} ) + E_{kn}^{i}( \mathbf{x}^{i} ) \right. \nonumber \\
    %+ E_{mv}^{i}( \mathbf{x}^{i} ) + E_{dg}^{i}( \mathbf{x}^{i} | \mathbf{o}^{i} ) \right\}.
    &\quad\quad\quad\quad\quad\quad\quad \left. + E_{mv}^{i}( \mathbf{x}^{i} ) + E_{dg}^{i}( \mathbf{x}^{i} | \mathbf{o}^{i} ) \right\}.
\label{eq:S_global_opt}
\end{align}
\section{Optimization of the Global Objective}
In the following, we introduce two optimization approaches to solve Eq.~\ref{eq:S_global_opt}. In general, our optimization framework is iterative in nature, like the algorithm described in Alg. 1 in the main manuscript with different options for optimizing Eq.~\ref{eq:S_global_opt}. First, we explain the message passing algorithm (MPA) and then we consider %interior point method (IPM) and finally 
the unscented Kalman smoother (UKS).
% \begin{algorithm}[t]
% 	% required keywords
%     \SetKwInOut{Input}{Input}
%     \SetKwInOut{Output}{Output}
% 	% keywords
% 	\SetKwData{Left}{left}
%     \SetKwData{This}{this}
%     \SetKwData{Up}{up}
% 	\SetKwFunction{Union}{Union}
%     \SetKwFunction{FindCompress}{FindCompress}
% 	% input/output
%     \Input{$\mathbf{O}$, $(\theta_{NN}, \sigma_{NN})$, $(\theta_{GP}, \mu_{GP}, \sigma_{GP})$, \\$C_{kn}$, $C_{mv}$}
%     \Output{$\hat{\mathbf{X}}$}
%     % algorithm body
%     Initialize $\mathbf{X}$\;
%     \Repeat(){$\mathbf{X}$ converges}{
%     	Compute DNN prior velocities of $\mathbf{X}$ using Eq.1 in the main manuscript\;
%         Compute GP prior velocities of $\mathbf{X}$ using Eq.3 in the main manuscript\;
%         Find $\mathbf{X}$ by minimizing Eq.~\ref{eq:global_opt}\;
%     }()
%     $\hat{\mathbf{X}} = \mathbf{X}$\;
%     % caption & label
% 	\caption{Proposed Optimization Framework}
%     \label{alg1}
% \end{algorithm}

%--------------------------------------------------------
\subsection{Message Passing Algorithm (MPA)}
The first optimization method, message-passing~\cite{bento13nips}, can be utilized to find $\mathbf{\hat{X}}$. In this algorithm, we consider each energy term that minimizes each point in a trajectory as an independent minimizer node, and the minimizer nodes are connected to one another via another type of nodes called equality nodes, building a bipartite graph. Minimizer nodes will minimize each variable, while equality nodes make sure that all minimizer nodes are up-to-date about dependent variables optimized by other minimizer nodes. Messages here will be agent location in different time step. The prior work~\cite{yoon16wacvws} had to link inter-agent minimizer nodes to deal with $E_{p}^{i,j}$. However, we only need to connect nodes within the same agent this time. 

More specifically, we first define messages from the equality nodes to the minimizer nodes as $\mathbf{n}_{t}^{i}$ for $i$-th agent at time step $t$. This message conveys the expected location of the agent predicted by the other agents in the previous iteration of the optimization. Then we denote $\mathbf{\hat{x}}_{t}^{i}$ as the optimized trajectory variable stored in the equality nodes, and $\mathbf{x}_{t}^{i}$ as the corresponding optimized trajectory variable stored in the minimizer node. A general strategy here, is to encode locally optimized $\mathbf{x}_{t}^{i}$ to be equal across all minimizers that are referencing $\mathbf{x}_{t}^{i}$ by iteratively passing errors in $\mathbf{x}_{t}^{i}$ estimates, and penalize the error when performing the optimization in the minimizer nodes. 

Then, how can we optimize each minimizer node? For $E_{gt}^{i}(\mathbf{x}_{t}^{i} | \mathbf{o}_{t}^{i})$, the term is a simple squared equation and since it is not dependent on any other nodes, we can solve it in closed form for the quadratic equation. For $E_{kn}(\mathbf{x}_{t}^{i}), E_{mv}(\mathbf{x}_{t}^{i})$, they are dependent on two variables, so we need two squared penalty terms to optimize, e.g., in case of $E_{kn}(\cdot)$, we need to solve
\begin{align}
\argmin_{\mathbf{x}_{t-1}^{i}, \mathbf{x}_{t}^{i}} E_{kn}^{i}(\cdot)
	&= \left[ C_{kn} \sum_{t} \| \mathbf{x}_{t}^{i} - \mathbf{x}_{t-1}^{i} \|^{2} \right. \nonumber \\
    &\left. + \frac{\rho}{2} \| \mathbf{x}_{t-1}^{i} - \mathbf{n}_{t-1}^{i} \|^{2} + \frac{\rho}{2} \| \mathbf{x}_{t}^{i} - \mathbf{n}_{t}^{i} \|^{2} \right],
\end{align}
where $\rho$ is the penalty weight. Original algorithm~\cite{bento13nips} uses dynamically changing weight value, but one can consider this as a fixed parameter, reducing the algorithm to the standard alternating direction of method of multipliers (ADMM). To solve this, we need to take first order derivative and solve for the variables $\mathbf{x}_{t-1}^{i}, \mathbf{x}_{t}^{i}$. In case of $E_{mv}(\cdot)$, we can utilize KKT conditions to find the closed form solution, as explained in~\cite{bento13nips}. It is worth noting that, in $E_{dg}^{i}(\cdot)$, from each minimizer's perspective, all other terms except the $\mathbf{x}_{t-1}^{i}, \mathbf{x}_{t}^{i}$ are constant in terms of optimization, so the minimizer solution should be identical to the $E_{kn}(\cdot)$ with minor difference in coefficients.

Specifically, Eq.~\ref{eq:S_new_unary_nonlin_loc} can be optimized as
\begin{align}
E_{dg}^{i}( \mathbf{x}^{i} | \mathbf{o}^{i} ) 
	&= \lambda \sum_{t} \left\Vert \mathbf{x}_{t}^{i} - \mathbf{x}_{t-1}^{i} - F \right\Vert^{2} \nonumber \\
    &\quad+ \frac{\rho}{2} \| \mathbf{x}_{t-1}^{i} - \mathbf{n}_{t-1}^{i} \|^{2} + \frac{\rho}{2} \| \mathbf{x}_{t}^{i} - \mathbf{n}_{t}^{i} \|^{2}
\end{align}
where $\lambda = 1 / (\sigma_{DNN}^2 \Delta t^{2})$ and $F$ to denote the nonlinear constant term. Taking derivative for each $t$-th term,
\begin{align}
\frac{\partial E_{dg}^{i}( \mathbf{x}_{t}^{i} | \mathbf{o}^{i} )}{\partial \mathbf{x}_{t-1}^{i}}
	&= -2 \lambda ( \mathbf{x}_{t}^{i} - \mathbf{x}_{t-1}^{i} - F ) + \rho ( \mathbf{x}_{t-1}^{i} - \mathbf{n}_{t-1}^{i} ) \\
\frac{\partial E_{dg}^{i}( \mathbf{x}_{t}^{i} | \mathbf{o}^{i} )}{\partial \mathbf{x}_{t}^{i}}
	&= 2 \lambda ( \mathbf{x}_{t}^{i} - \mathbf{x}_{t-1}^{i} - F ) + \rho ( \mathbf{x}_{t}^{i} - \mathbf{n}_{t}^{i} ),
\end{align}
Then, by setting the above as zero and solving for $\mathbf{x}_{t}^{i}$ and $\mathbf{x}_{t-1}^{i}$, we get
\begin{align}
0 
	&= -2 \lambda ( \mathbf{x}_{t}^{i} - \mathbf{x}_{t-1}^{i} - F ) + \rho ( \mathbf{x}_{t-1}^{i} - \mathbf{n}_{t-1}^{i} ), \\
0 
	&= 2 \lambda ( \mathbf{x}_{t}^{i} - \mathbf{x}_{t-1}^{i} - F ) + \rho ( \mathbf{x}_{t}^{i} - \mathbf{n}_{t}^{i} ), \\
\mathbf{x}_{t-1}^{i} 
	&= \frac{(2 \lambda + \rho) \rho \mathbf{n}_{t-1}^{i} + 2 \lambda \rho \mathbf{n}_{t}^{i} + 2 \lambda F (2 \lambda + \rho - 1)}{(4 \lambda + \rho) \rho} \\
\mathbf{x}_{t}^{i} 
	&= \frac{(2 \lambda + \rho) \rho \mathbf{n}_{t}^{i} + 2 \lambda \rho \mathbf{n}_{t-1}^{i} + 2 \lambda F (2 \lambda + \rho + 1)}{(4 \lambda + \rho) \rho}
\end{align}

%--------------------------------------------------------
% \subsection{Interior Point Method (IPM)}
% The second optimization method to solve Eq.~\ref{eq:global_opt}, interior point method, is a generic constrained optimization solver that appends to the original objective function a scaled penalty function which is called a barrier function~\cite{boyd2004convex} with a scaled factor named barrier parameter. The barrier function maps a solution point to a penalty value, and has the property that the penalty approaches to infinity when the solution point approaches to the boundary of the feasible region. In this way, a solution point is dissuaded from exiting the feasible region due to the high penalty and thus the original constrained optimization problem is converted into an unconstrained optimization problem. The barrier convergence theorem shows that when the barrier parameter approaches to zero, the solution of the unconstrained barrier objective approaches to the solution of the original problem. For numerical stability, the actual computational procedure of interior point is an iterative procedure of the primal-dual update using Newton's method.

%--------------------------------------------------------
\subsection{Unscented Kalman Smoother (UKS)}
In this subsection, we approximate the constrained optimization problem Eq.~\ref{eq:S_global_opt} with a dynamical system, and solve it efficiently using unscented Kalman smoother. First, consider the following minimization problem:

\begin{align}
\label{eq:S_obj_uks}
J(\mathbf{x}) &= \frac{1}{2}(\mathbf{x} - \mathbf{a})^{\top} \mathbf{A} (\mathbf{x} - \mathbf{a}) + \frac{1}{2}(\mathbf{x} - \mathbf{b})^{\top} \mathbf{B} (\mathbf{x} - \mathbf{b}) \\
s.t.&\quad \| \mathbf{v} \| \le C_{mv}
\end{align}
where
\begin{align}
\label{eq:S_tmp1}
\mathbf{x} &= \mathbf{x}_{t+1} \\
\label{eq:S_tmp2}
\mathbf{a} &= \mathbf{x}_{t} \\
\label{eq:S_tmp3}
\mathbf{A} &= 2 C_{kn} \mathbf{I} \\
\label{eq:S_tmp4}
\mathbf{b} &= \mathbf{x}_{t} + \Delta t \cdot f_{NN-GP}(\mathbf{o}_{t}) \\
\label{eq:S_tmp5}
\mathbf{B} &= 2 \lambda \mathbf{I}.
\end{align}
Note that we suppressed the superscript $i$ since we will optimize each agent's trajectory independently. 
Substituting Eq.~\ref{eq:S_tmp1}-Eq.\ref{eq:S_tmp5} back into Eq.~\ref{eq:S_obj_uks}, it is clear that 
\begin{align}
J(\mathbf{x}_{t+1}) &= C_{kn}|| \mathbf{x}_{t+1} - \mathbf{x}_{t} ||^2 \nonumber \\
&\quad+ \lambda|| \mathbf{x}_{t+1} - \mathbf{x}_{t} - \Delta t \cdot f_{NN-GP}(\mathbf{o}_{t})||^2, 
\end{align}
which is the summation of the kinetic energy term and the newly introduce prior energy term. If we ignore the velocity constraint $\quad \| \mathbf{v} \| \le C_{mv}$, setting the first derivative of Eq.~\ref{eq:S_obj_uks} to zero immediately gives the solution of the minimization problem $\mathbf{x} =(\mathbf{A}+\mathbf{B})^{-1}(\mathbf{A}\mathbf{a}+\mathbf{B}\mathbf{b})$. Substituting Eq.~\ref{eq:S_tmp1}-Eq.\ref{eq:S_tmp5} back into the solution, we get: 
\begin{align}
\mathbf{x}_{t+1} & = \mathbf{x}_{t} + \Delta t \cdot \frac{\lambda}{\lambda + C_{kn}} \cdot f_{NN-GP}(\mathbf{o}_{t}) \nonumber \\ 
        & \approx \mathbf{x}_{t} + \Delta t \cdot f_{NN-GP}(\mathbf{o}_{t}) + \epsilon \nonumber \\
        & = \mathcal{T}(\mathbf{x}_{t}),
\label{eq:S_transition_dynamical_system}
\end{align}
where 
\begin{align}
\bm{\epsilon} \sim \mathcal{N}(\mathbf{0}, (\mathbf{A}+\mathbf{B})^{-1}) = \mathcal{N}(\mathbf{0}, \frac{1}{ 2( C_{kn}+\lambda ) }\mathbf{I}).
\end{align}

The interpretation of Eq.~\ref{eq:S_transition_dynamical_system} is that the true location of an agent $\mathbf{x}_{t+1}$ at time $t+1$, starting from the true location $\mathbf{x}_{t}$ at time $t$, is determined by the average velocity from a prior velocity model for a time period $\Delta t$, with error $\bm{\epsilon}$. The error $\bm{\epsilon}$ measures the uncertainty of the prior velocity model about the unknown, true velocity. Thus the error $\bm{\epsilon}$ could be modeled as a Gaussian noise. In this way, Eq.~\ref{eq:S_transition_dynamical_system} implies a transition function from $\mathbf{x}_{t}$ to $\mathbf{x}_{t+1}$ of a dynamical system, and $\bm{\epsilon}$ denotes the process noise, which is Gaussian.

Next, let us consider the constraint of the above minimization problem. The velocity in this case is `noisy', expressed as $\mathbf{v} = f_{NN-GP}(\mathbf{o}_{t}) + \frac{1}{\Delta t}\bm{\epsilon}$. One can show that, if we directly pass the 2-dimensional velocity $\mathbf{v}$ to a hard magnitude limiter
\begin{align}
f(\mathbf{v}) = 
\begin{cases}
\mathbf{v}	&\text{if } \| \mathbf{v} \| \leq C_{mv} \\
C_{mv} 		&\text{otherwise}
\end{cases},
\end{align}
%$f(v) = v$ if $||v|| \leq C_{mv}$ otherwise $f(v) = C_{mv}$, 
the dynamical system will become explosively complex along transitions. Instead, we apply the magnitude limiter to each of the velocity component as $|v_{x}| < C_{mv}$, $|v_{y}| < C_{mv}$ with a soft magnitude limiter: $f(v) = 2 C_{mv} \cdot (s(v) -1/2)$ where $v$ is either $v_{x}$ or $v_{y}$ and $s(\cdot)$ the sigmoid function. 

The key point is that we do not directly compute $f(\mathbf{v})$ given $\mathbf{v}$. Instead, first-order Taylor approximation is considered: 
\begin{align}
f(v) \approx f(a) + f'(a)(v - a)
\label{eq:S_first-order_Taylor_approx}
\end{align}
where $a$ is a constant (a is a moving point such that it is close to the current $v$) given $v$. Therefore, $f(v)$ is a linear transformation of $v$, hence Gaussian if $v$ is Gaussian. This recursive property makes the dynamical system always a Gaussian, avoiding being explosively complex along transitions. Thus, if we use $[f(v_{x}), f(v_{y})]^{T}$ rather than $\mathbf{v}$ to conduct the transition in Eq.~\ref{eq:S_transition_dynamical_system}, the maximum-velocity constraint is applied.

Furthermore, if the measurement function of the dynamical system is defined as:
\begin{align}
\mathbf{o}_{t} = \mathbf{x}_{t} + \mathbf{r}_{t} = H(\mathbf{x}_{t}),
\label{eq:S_measurement_dynamical_system}
\end{align}
where the Gaussian measurement noise $\mathbf{r}_{t}$ adaptively controls the tracker output energy term $E_{gt}(\mathbf{x}_{t}, \mathbf{o}_{t}) = u_{t}||\mathbf{x}_{t} - \mathbf{o}_{t}||^2 = u_{t}||\mathbf{r}_{t}||^2$. For instance, we know that which points are visible and which points are missing given the incomplete trajectories. If $u_{t}=1$, denoting that $\mathbf{o}_{t}$ is an actually observed point, the covariance of $\mathbf{r}_{t}$ could be set as the identity matrix scaled by a small number, indicating high confidence and hence low energy. On the other hand, if $u_{t}=0$, denoting that $\mathbf{o}_{t}$ is only from inference, the covariance of $\mathbf{r}_{t}$ would be large when estimated from training data, indicating low confidence and high energy.

Overall, applying unscented Kalman filter followed by Rauch-Tung-Striebel smoother~\cite{Rauch1965} that calculates "smoothed" sequence from the given filter output optimizes the approximated objective accumulated over all time steps.

%--------------------------------------------------------
\section{Experiments}

\begin{figure}[t]
\centering
\includegraphics[width = 0.38\textwidth]{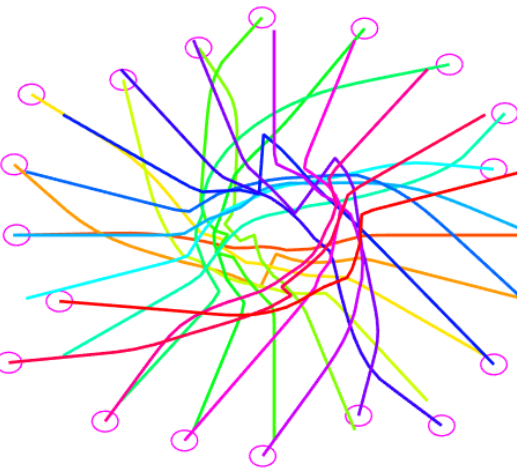} \\
\includegraphics[width = 0.38\textwidth]{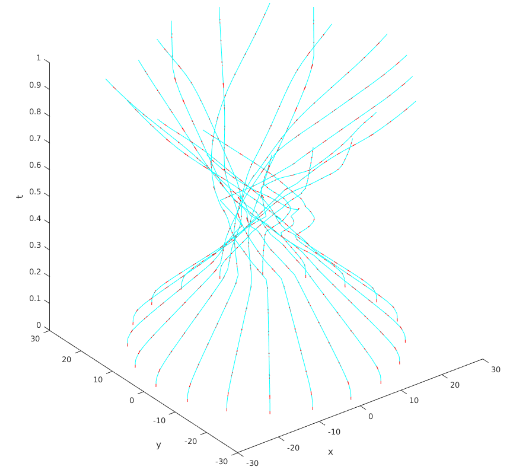}
\caption{%Top row shows Zoomed in images of Fig.~\ref{fig:visualization1}. Bottom row 
The optimized trajectories in concentric-circles scenario in (top) 2D view and (bottom) normalized time axis. Note the rotating behavior of agents in this scenario.}
\label{fig:S_visualization2}
\end{figure}

To evaluate the proposed framework for various combinations of priors and optimization methods, we prepare similar experimental settings as in~\cite{yoon16wacvws}. We prepare 6 scenarios (environmental configurations): 3 different settings of bottlenecks (each contains a challenging exit to go through for evacuation), concentric circle (agents are symmetrically placed along a circle and aim to reach their antipodal positions), two-way and four-way hallways (the environment is divided by two or four building blocks and agents move along the regulated ways). These scenarios were previously introduced in~\cite{steerbench} and are representative settings for studying crowd behaviors. For each scenario, we simulate 3,000 frames of 30-40 agents except for the concentric circle scenario, where there are 20 agents, followed by sub-sampling those frames. Detailed statistics of the scenarios are summarized in Table~\ref{tab:S_scenario_list}. An example visualization of the scenarios is shown in Fig.~\ref{fig:S_visualization2}. The ground truth trajectories were obtained by randomly setting the agents' initial locations and running SteerSuite~\cite{steersuite2009} library driven by social force AI~\cite{PhysRevE.51.4282}. Then the trajectories were split into training set and testing set with 6:1 ratio.

GP regressor is trained per-scenario. More specifically, for each scenario, we use 1,000 spatio-temporal points which are three dimensional denoted by $(x,y,t)$ to fit one GP regression model for the x component of the velocity, and use the same 1,000 spatio-temporal points to fit the second GP regression model for the y component of the velocity. We treat the two velocity components as two independent Gaussian processes and the final velocity is simply their superposition. We conduct data normalization and auto relevance determination, and choose the Matern 5/2 kernel. Note that even though there are distinct destinations of trajectories in the two-way hallway scenario and in the four-way hallway, one set of GPs are sufficient to model multiple modalities of movements in a single scenario.

NN regressor is trained using around 800,000 samples across all 6 scenarios, and tested in test simulations involving around 130,000 test samples. There are several branches in NN and all of them are fully connected followed by ReLU activations and dropout layers. If we adopt the linear combination of NN and GP, there are three input branches for NN: 360 dimensional distance map, $360\times 2$ dimensional velocity map and 2 dimensional global desired velocity. For GP-fed NN, there are two additional branches for NN: 2 dimensional GP local guidance velocity and 2 dimensional GP standard deviation. These branches merge at a higher level (at the 6-th layer), which is still fully connected, with ReLU activation and dropout layers. The output layer is a 2 dimensional dense layer for velocity regression. The width of the network is 1,024 while the depth is 10. Mean squared error is the loss function and RMSprop is the learner. With early stop, we set learning rate to be 0.0001, batch size 64 and dropout rate 0.2. In addition, similar to the work in~\cite{long2017}, we also apply a clamping to regulate the output of the NN.

%---------------------------------------------------------------------------------
\begin{table}
\centering
\caption{List of scenarios we have tested and the environment width and height in each scenario (in meters), and number of frames subsampled.}
\label{tab:S_scenario_list}
\begin{tabular}{l|c|r}
\toprule
Scenario				& Width	$\times$ Height & \#frames\\
\midrule
bottleneck-evacuation 	& $200 \times 160$ 	&	101 \\
bottleneck-evacuation2 	& $100 \times 80$ 	&	53	\\
bottleneck-squeeze 		& $200 \times 200$ 	&	94	\\
concentric-circles 		& $20 \times 20$ 	&	37	\\
hallway-two-way 		& $200 \times 200$ 	&	101	\\
hallway-four-way 		& $200 \times 200$ 	&	101	\\
\bottomrule
\end{tabular}
\end{table}

\bibliographystyle{aaai}
\bibliography{sigproc,mypubs,sejong} 

\begin{thebibliography}{}

\bibitem[\protect\citeauthoryear{Alahi, Ramanathan, and
  Fei-Fei}{2014}]{alahi2014}
Alahi, A.; Ramanathan, V.; and Fei-Fei, L.
\newblock 2014.
\newblock Socially-aware large-scale crowd forecasting.
\newblock In {\em Computer Vision and Pattern Recognition (CVPR), 2014 IEEE
  Conference on},  2211--2218.

\bibitem[\protect\citeauthoryear{Ali and Shah}{2008}]{ali2008}
Ali, S., and Shah, M.
\newblock 2008.
\newblock {Floor Fields for Tracking in High Density Crowd Scenes}.
\newblock In {\em ECCV}.

\bibitem[\protect\citeauthoryear{Ali \bgroup et al\mbox.\egroup
  }{2013}]{Ali:2013:MSV:2568128}
Ali, S.; Nishino, K.; Manocha, D.; and Shah, M.
\newblock 2013.
\newblock {\em Modeling, Simulation and Visual Analysis of Crowds: A
  Multidisciplinary Perspective}.
\newblock Springer Publishing Company, Incorporated.

\bibitem[\protect\citeauthoryear{Bento \bgroup et al\mbox.\egroup
  }{2013}]{bento13nips}
Bento, J.; Derbinsky, N.; Alonso-Mora, J.; and Yedidia, J.~S.
\newblock 2013.
\newblock A message-passing algorithm for multi-agent trajectory planning.
\newblock In Burges, C. J.~C.; Bottou, L.; Welling, M.; Ghahramani, Z.; and
  Weinberger, K.~Q., eds., {\em Advances in Neural Information Processing
  Systems 26}. Curran Associates, Inc.
\newblock  521--529.

\bibitem[\protect\citeauthoryear{Bera and Manocha}{2014}]{bera2014}
Bera, A., and Manocha, D.
\newblock 2014.
\newblock {Realtime Multilevel Crowd Tracking using Reciprocal Velocity
  Obstacles}.
\newblock In {\em 22nd International Conference on Pattern Recognition}.

\bibitem[\protect\citeauthoryear{Bera, Kim, and Manocha}{2015}]{bera2015}
Bera, A.; Kim, S.; and Manocha, D.
\newblock 2015.
\newblock {Efficient Trajectory Extraction and Parameter Learning for
  Data-Driven Crowd Simulation}.
\newblock In {\em Graphics Interface}.

\bibitem[\protect\citeauthoryear{Bera, Kim, and Manocha}{2016}]{Bera2016}
Bera, A.; Kim, S.; and Manocha, D.
\newblock 2016.
\newblock Online parameter learning for data-driven crowd simulation and
  content generation.
\newblock {\em Computers and Graphics}.

\bibitem[\protect\citeauthoryear{Helbing and Moln\'ar}{1995}]{PhysRevE.51.4282}
Helbing, D., and Moln\'ar, P.
\newblock 1995.
\newblock Social force model for pedestrian dynamics.
\newblock {\em Phys. Rev. E} 51:4282--4286.

\bibitem[\protect\citeauthoryear{Hu, Ali, and Shah}{2008}]{hu2008}
Hu, M.; Ali, S.; and Shah, M.
\newblock 2008.
\newblock {Learning Motion Patterns in Crowded Scenes Using Motion Flow Field}.
\newblock In {\em ICPR}.

\bibitem[\protect\citeauthoryear{Jacques~Junior, Raupp~Musse, and
  Jung}{2010}]{musse2010}
Jacques~Junior, J.; Raupp~Musse, S.; and Jung, C.
\newblock 2010.
\newblock Crowd analysis using computer vision techniques.
\newblock {\em Signal Processing Magazine, IEEE} 27(5):66--77.

\bibitem[\protect\citeauthoryear{Kapadia, Pelechano, and
  Allbeck}{2015}]{kapadia2015virtual}
Kapadia, M.; Pelechano, N.; and Allbeck, J.
\newblock 2015.
\newblock {\em Virtual Crowds: Steps Toward Behavioral Realism}.
\newblock MORGAN \& CLAYPOOL.

\bibitem[\protect\citeauthoryear{Kim, Lee, and Essa}{2011}]{Kim2011-zf}
Kim, K.; Lee, D.; and Essa, I.
\newblock 2011.
\newblock {Gaussian Process Regression Flow for Analysis of Motion
  Trajectories}.
\newblock In {\em IEEE International Conference on Computer Vision, (ICCV)}.

\bibitem[\protect\citeauthoryear{Lerner, Chrysanthou, and
  Lischinski}{2007}]{journals/cgf/LernerCL07}
Lerner, A.; Chrysanthou, Y.; and Lischinski, D.
\newblock 2007.
\newblock Crowds by example.
\newblock {\em Comput. Graph. Forum} 26(3):655--664.

\bibitem[\protect\citeauthoryear{Li \bgroup et al\mbox.\egroup }{2015}]{li2015}
Li, T.; Chang, H.; Wang, M.; Ni, B.; Hong, R.; and Yan, S.
\newblock 2015.
\newblock Crowded scene analysis: A survey.
\newblock {\em IEEE Transactions on Circuits and Systems for Video Technology}
  25(3):367--386.

\bibitem[\protect\citeauthoryear{Long, Liu, and Pan}{2017}]{long2017}
Long, P.; Liu, W.; and Pan, J.
\newblock 2017.
\newblock Deep-learned collision avoidance policy for distributed multiagent
  navigation.
\newblock {\em IEEE Robotics and Automation Letters} 2(2):656--663.

\bibitem[\protect\citeauthoryear{Pellegrini and Gool}{2013}]{eth_biwi_01014}
Pellegrini, S., and Gool, L.~V.
\newblock 2013.
\newblock Tracking with a mixed continuous-discrete conditional random field.
\newblock {\em Computer Vision and Image Understanding}.

\bibitem[\protect\citeauthoryear{Pellegrini \bgroup et al\mbox.\egroup
  }{2009}]{pellegrini2009}
Pellegrini, S.; Ess, A.; Schindler, K.; and van Gool, L.
\newblock 2009.
\newblock {You'll Never Walk Alone: Modeling Social Behavior for Multi-target
  Tracking}.
\newblock In {\em IEEE International Conference on Computer Vision}.

\bibitem[\protect\citeauthoryear{Pellegrini \bgroup et al\mbox.\egroup
  }{2010}]{eth_biwi_00785}
Pellegrini, S.; Ess, A.; Tanaskovic, M.; and Gool, L.~V.
\newblock 2010.
\newblock Wrong turn - no dead end: a stochastic pedestrian motion model.
\newblock In {\em International Workshop on Socially Intelligent Surveillance
  and Monitoring (SISM)}.

\bibitem[\protect\citeauthoryear{Rauch, Striebel, and Tung}{1965}]{Rauch1965}
Rauch, H.; Striebel, C.; and Tung, F.
\newblock 1965.
\newblock Maximum likelihood estimates of linear dynamic systems.
\newblock {\em AIAA Journal} 3(8):1445--1450.

\bibitem[\protect\citeauthoryear{Rodriguez, Ali, and
  Kanade}{2009}]{rodriguez2009}
Rodriguez, M.; Ali, S.; and Kanade, T.
\newblock 2009.
\newblock {Tracking in Unstructured Crowded Scenes}.
\newblock In {\em ICCV}.

\bibitem[\protect\citeauthoryear{Rodriguez \bgroup et al\mbox.\egroup
  }{2011}]{Rodriguez2011}
Rodriguez, M.; Sivic, J.; Laptev, I.; and Audibert, J.-Y.
\newblock 2011.
\newblock {Data-driven Crowd Analysis in Videos}.
\newblock In {\em IEEE International Conference on Computer Vision}.

\bibitem[\protect\citeauthoryear{Sharma, Huang, and Nevatia}{2012}]{sharma2012}
Sharma, P.; Huang, C.; and Nevatia, R.
\newblock 2012.
\newblock Unsupervised incremental learning for improved object detection in a
  video.
\newblock In {\em 2012 IEEE Conference on Computer Vision and Pattern
  Recognition},  3298--3305.

\bibitem[\protect\citeauthoryear{Singh \bgroup et al\mbox.\egroup
  }{2009a}]{steersuite2009}
Singh, S.; Kapadia, M.; Faloutsos, P.; and Reinman, G.
\newblock 2009a.
\newblock {An Open Framework for Developing, Evaluating, and Sharing Steering
  Algorithms}.
\newblock In {\em Proceedings of the 2nd International Workshop on Motion in
  Games}, MIG '09,  158--169.
\newblock Berlin, Heidelberg: Springer-Verlag.

\bibitem[\protect\citeauthoryear{Singh \bgroup et al\mbox.\egroup
  }{2009b}]{steerbench}
Singh, S.; Kapadia, M.; Faloutsos, P.; and Reinman, G.
\newblock 2009b.
\newblock {SteerBench: a benchmark suite for evaluating steering behaviors}.
\newblock {\em Computer Animation and Virtual Worlds} 9999(9999):n/a+.

\bibitem[\protect\citeauthoryear{van~den Berg \bgroup et al\mbox.\egroup
  }{2011}]{berg2011}
van~den Berg, J.; Guy, S.~J.; Lin, M.~C.; and Manocha, D.
\newblock 2011.
\newblock {Reciprocal n-Body Collision Avoidance}.
\newblock In {\em International Symposium on Robotic Research (ISRR)}.

\bibitem[\protect\citeauthoryear{van~den Berg, Lin, and
  Manocha}{2008}]{berg2008}
van~den Berg, J.; Lin, M.~C.; and Manocha, D.
\newblock 2008.
\newblock {Reciprocal Velocity Obstacles for Real-Time Multi-Agent Navigation}.
\newblock In {\em Proceedings of the IEEE International Conference on Robotics
  and Automation (ICRA)}.

\bibitem[\protect\citeauthoryear{Yoon \bgroup et al\mbox.\egroup
  }{2016}]{yoon16wacvws}
Yoon, S.; Kapadia, M.; Sahu, P.; and Pavlovic, V.
\newblock 2016.
\newblock Filling in the blanks: Reconstructing microscopic crowd motion from
  multiple disparate noisy sensors.
\newblock In {\em 2016 {IEEE} Winter Applications of Computer Vision Workshops,
  {WACV} 2016 Workshops},  1--9.
\newblock 25\% contribution.

\bibitem[\protect\citeauthoryear{Zhan \bgroup et al\mbox.\egroup
  }{2008}]{zhan2008}
Zhan, B.; Monekosso, D.; Remagnino, P.; Velastin, S.; and Xu, L.-Q.
\newblock 2008.
\newblock {Crowd analysis: a survey}.
\newblock {\em Machine Vision Applications} 19:345?--357.

\end{thebibliography}

\end{document}